\documentclass[lettersize,journal]{IEEEtran}
\usepackage{amsmath,amsfonts}
\usepackage{algorithmic}
\usepackage{algorithm}
\usepackage{array}
\usepackage[caption=false,font=normalsize,labelfont=sf,textfont=sf]{subfig}
\usepackage{textcomp}
\usepackage{stfloats}
\usepackage{url}
\usepackage{verbatim}
\usepackage{graphicx}
\usepackage{cite}
\usepackage{bm}
\usepackage{amsmath}
\usepackage{amssymb}

\usepackage{pifont}
\usepackage{makecell}
\newcommand{\tabincell}[2]{\begin{tabular}
{@{}#1@{}}#2\end{tabular}}
\newcommand{\xmark}{\ding{55}}

\usepackage{color}

\usepackage{multirow}
\usepackage{hhline}
\begin{document}
\bstctlcite{IEEEexample:BSTcontrol}

\title{Confidence Score Based Speaker Adaptation of Conformer Speech Recognition Systems}

\author{Jiajun Deng, Xurong Xie, Tianzi Wang, Mingyu Cui, Boyang Xue, Zengrui Jin, \\ Guinan Li, Shujie Hu, Xunying Liu,~\IEEEmembership{Member,~IEEE}
\thanks{Jiajun Deng, Tianzi Wang, Mingyu Cui, Boyang Xue, Zengrui Jin, Guinan Li, Shujie Hu are with the Chinese University of Hong Kong, China (email: \{jjdeng, twang, mycui, byxue, zrjin, gnli, sjhu\}@se.cuhk.edu.hk)}
\thanks{Xurong Xie is with Institute of Software, Chinese Academy of Sciences, Beijing, China (email: xurong@iscas.ac.cn).}
\thanks{Xunying Liu is with the Chinese University of Hong Kong, China (email:  xyliu@se.cuhk.edu.hk).} 
\thanks{Corresponding authors: Xurong Xie and Xunying Liu.}
}

\markboth{Journal of \LaTeX\ Class Files,~Vol.~14, No.~8, August~2021}%
{Shell \MakeLowercase{\textit{et al.}}: A Sample Article Using IEEEtran.cls for IEEE Journals}

\maketitle

\begin{abstract}
Speaker adaptation techniques provide a powerful solution to customise automatic speech recognition (ASR) systems for individual users. Practical application of unsupervised model-based speaker adaptation techniques to data intensive end-to-end ASR systems is hindered by the scarcity of speaker-level data and performance sensitivity to transcription errors. To address these issues, a set of compact and data efficient speaker-dependent (SD) parameter representations are used to facilitate both speaker adaptive training and test-time unsupervised speaker adaptation of state-of-the-art Conformer ASR systems. The sensitivity to supervision quality is reduced using a confidence score-based selection of the less erroneous subset of speaker-level adaptation data. Two lightweight confidence score estimation modules are proposed to produce more reliable confidence scores. The data sparsity issue, which is exacerbated by data selection, is addressed by modelling the SD parameter uncertainty using Bayesian learning. Experiments on the benchmark 300-hour Switchboard and the 233-hour AMI datasets suggest that the proposed confidence score-based adaptation schemes consistently outperformed the baseline speaker-independent (SI) Conformer model and conventional non-Bayesian, point estimate-based adaptation using no speaker data selection. Similar consistent performance improvements were retained after external Transformer and LSTM language model rescoring. In particular, on the 300-hour Switchboard corpus, statistically significant WER reductions of 1.0\%, 1.3\%, and 1.4\% absolute (9.5\%, 10.9\%, and 11.3\% relative) were obtained over the baseline SI Conformer on the NIST Hub5’00, RT02, and RT03 evaluation sets respectively. Similar WER reductions of 2.7\% and 3.3\% absolute (8.9\% and 10.2\% relative) were also obtained on the AMI development and evaluation sets. 
\end{abstract}
\vspace{-0.1cm}

\vspace{-0.2cm}
\section{Introduction}
\label{sec:intro}
\IEEEPARstart{T}{he} performance of automatic speech recognition (ASR) systems has been significantly improved over the past decades with the wide application of deep learning techniques
\cite{dahl2011context,abdel2014convolutional,peddinti2015time,chan2016listen,watanabe2017hybrid,Li2020OnTC,dong2018speech,nakatani2019improving,bell2020adaptation,wang2020transformer,gulati2020conformer,guo2021recent,li2022recent,saon2021advancing,deng2022confidence}. More recently there has been a major trend in the speech technology field transiting from hybrid ASR systems\cite{dahl2011context,peddinti2015time,abdel2014convolutional} to end-to-end (E2E) modelling \cite{chan2016listen,watanabe2017hybrid,dong2018speech,nakatani2019improving,wang2020transformer,gulati2020conformer,Li2020OnTC} which utilizes a single neural network to directly map acoustic feature vectors to the surface word or token sequences. E2E ASR systems represented by connectionist temporal classification (CTC) \cite{graves2006connectionist}, listen, attend and spell (LAS) \cite{chan2016listen} or the attention-based encoder-decoder (AED) \cite{chorowski2015attention}, recurrent neural network transducer (RNN-T) model \cite{graves2012sequence,rao2017exploring}, and Transformer \cite{dong2018speech,Karita2019ACS} or convolution-augmented Transformer (Conformer) \cite{gulati2020conformer}, are optimized using a single objective function which is consistent with speech recognition error rate. Among these E2E models, Conformer encoder structures leverage a combined use of self-attention and convolution modules that are designed to learn both the longer-range, global contexts and local acoustic features. These advantages also lead to their wide adoption in both Conformer-transformer AED and Conformer transducers \cite{gulati2020conformer,guo2021recent,li2022recent,tuske2021limit,huang2021rapid,zeineldeen2022improving}.

A key challenge for ASR systems, including those based on Conformers, is to model the systematic and latent variation in speech. A major source of such variability is attributable to speaker-level characteristics representing factors such as accent and idiosyncrasy, or physiological differences manifested in, for instance, age or gender. The diversity among speakers often creates a significant mismatch between the training and evaluation data distributions, leading to the degradation of recognition performance. To this end, speaker adaptation techniques, aiming to reduce the mismatch arising from the acoustic variability across speakers, play a vital role in current speech recognition systems. Separate reviews of speaker adaptation for conventional HMM, Hybrid ASR systems and E2E ASR systems are presented in Sec. \ref{subsection_1_1} and \ref{subsection_1_2}. 

\vspace{-0.3cm}
\subsection{Speaker Adaptation for HMM and Hybrid ASR Systems} \label{subsection_1_1}
Speaker adaptation techniques designed for HMM and hybrid ASR systems can be characterized into three general categories: \textbf{1)} Auxiliary speaker embedding-based approaches \cite{abdel2013fast,karafiat2011ivector,saon2013speaker,huang2015investigation}. \textbf{2)} Feature transformation-based methods \cite{digalakis1995speaker,gales1998maximum,eide1996parametric,lee1996speaker,uebel1999investigation,parthasarathi2015fmllr}. 
\textbf{3)} Model-based adaptation approaches \cite{liao2013speaker,yu2013kl,huang2020acoustic,xue2014fast,gales2000cluster,wu2015multi,tan2015cluster,Swietojanski2014LearningHU,Swietojanski2016LearningHU,xie2019blhuc,Xie2021BayesianLF,Zhang2016DNNSA,Neto1995SpeakeradaptationFH,Gemello2007LinearHT,Li2010ComparisonOD,Zhao2016LowrankPD,anastasakos1996compact,ochiai2014speaker,kitza2018comparison,meng2018speaker,tsuchiya2018speaker,price2014speaker,swietojanski2015structured,huang2015rapid,huang2015maximum,Huang2017BayesianUB}.

In the auxiliary speaker embedding-based approaches, acoustic features are augmented with the compact speaker-dependent (SD) embedding vectors computed for each speaker to facilitate speaker adaptation during both ASR system training and evaluation. The auxiliary speaker embedding vectors can be estimated either independently from the back-end recognition systems, for example, i-vectors \cite{karafiat2011ivector,saon2013speaker} that are learned from Gaussian mixed model (GMM) based universal background models, or jointly with the parameters of acoustic models, for example, speaker codes \cite{abdel2013fast} that are estimated via an alternating update between the auxiliary speaker embedding vectors and the parameters of back-end acoustic models. 

In the feature transformation-based methods, feature-space transformations are applied to acoustic front-ends to produce speaker-invariant input features that are then fed into the back-end recognition system. A widely used method for estimating feature-space transforms is feature-space maximum likelihood linear regression (f-MLLR) \cite{gales1998maximum}. f-MLLR transforms are estimated at speaker-level from GMM-HMM based systems \cite{digalakis1995speaker,gales1998maximum} before being applied to the acoustic front-ends to normalize the speaker level variability \cite{parthasarathi2015fmllr}. To compensate for the vocal tract length diversity among speakers, physiologically motivated vocal tract length normalization (VTLN) based feature transformations can also be used to produce speaker-invariant-like features \cite{lee1996speaker,eide1996parametric,uebel1999investigation}. Speaker-level VTLN normalized features can be obtained by applying the piece-wise linear frequency warping factors \cite{lee1996speaker, uebel1999investigation} to normalize the spectral features. 

In model-based adaptation approaches, the SD modelling granularity, varying between updating the whole and a small subset of acoustic model parameters \cite{liao2013speaker}, crucially depends on the precise amount of target-speaker adaptation data available. To ensure good generalization and prevent over-fitting to limited adaptation data in practical applications, separately designed compact forms of SD parameters based on structured transforms are often incorporated into various neural network layers. These include the use of linear input network (LIN) \cite{Neto1995SpeakeradaptationFH,Li2010ComparisonOD}, linear hidden network (LHN) \cite{Gemello2007LinearHT}, linear output network (LON) \cite{Gemello2007LinearHT}, learning hidden unit contributions (LHUC) \cite{Swietojanski2014LearningHU,Swietojanski2016LearningHU,xie2019blhuc,Xie2021BayesianLF}, parameterized activation function (PAct) \cite{Zhang2016DNNSA}, factorized linear transformation \cite{Zhao2016LowrankPD}, speaker code-based bias vector \cite{xue2014fast}, and adaptive linear interpolation of speaker-independent (SI) hidden layer output or parameter bases \cite{gales2000cluster,wu2015multi,tan2015cluster}. Alternatively, speaker adaptive training (SAT) \cite{anastasakos1996compact,ochiai2014speaker} allowing a joint optimization of both the SD and SI parameters at the training stage, can be further integrated with the above structured transform-based adaptations to provide a canonical model that can be better adapted to unseen speakers at test-time adaptation. To further ensure robustness of SD parameter estimation, regularization of the speaker adaptation objective function, including the use of L2 \cite{kitza2018comparison} or Kullback–Leibler (KL) divergence regularization \cite{liao2013speaker,yu2013kl,huang2020acoustic}, maximum a posterior adaptation \cite{huang2015maximum,Huang2017BayesianUB}, and variational inference-based Bayesian adaptation that models the uncertainty of SD parameters \cite{xie2019blhuc,Xie2021BayesianLF} have been studied for hybrid ASR systems. Furthermore, alternative speaker adaptation objective functions, including the use of adversarial learning \cite{meng2018speaker,tsuchiya2018speaker} and multi-task learning \cite{price2014speaker,swietojanski2015structured,huang2015rapid} have also been investigated. 

\vspace{-0.1cm}
\subsection{Speaker Adaptation for E2E ASR Systems}
\label{subsection_1_2}
In contrast, limited previous speaker adaptation research has been conducted for E2E ASR systems. Among these, auxiliary speaker adaptive features based on i-vector \cite{tuske2021limit,zeineldeen2022improving}, x-vector \cite{zeineldeen2022improving,baskar2022speaker} and f-MLLR \cite{baskar2022speaker}, or extracted from sequence summary network \cite{delcroix2018auxiliary}, attention-based speaker memory \cite{sari2020unsupervised} and speaker-aware modules \cite{fan2019speaker,zhao2020speech} are incorporated into attention-based encoder-decoder, or conventional, non-convolution augmented Transformer models. Several model-based speaker adaptation methods for E2E models utilize speaker adaptive neural network internal components, for example, SD neural beamforming, encoder, attention or decoder modules in multichannel E2E systems \cite{ochiai2018speaker}, while keeping the parameters of other components fixed as speaker independent during adaptation. Similar approaches were also developed for RNN-T \cite{sim2021robust,huang2020rapid} and Conformer transducer models \cite{huang2021rapid}. To mitigate the risk of overfitting to limited speaker-level data during adaptation, KL divergence regularization \cite{li2018speaker,meng2019speaker,weninger2019listen} and target-speaker data augmentation by text-to-speech synthesis \cite{sim2019personalization,huang2020rapid} approaches were proposed. Multi-task learning \cite{li2018speaker,meng2019speaker} and adversarial learning \cite{meng2019speaker} have been applied to CTC-based or attention-based E2E models. 

Efforts to develop model-based speaker adaptation techniques for Conformer models face several challenges. First, the often limited amounts of speaker-specific data require highly compact SD parameter representations to be used to mitigate the risk of over-fitting during adaptation. Second, SD parameter estimation is sensitive to the underlying supervision error rate \cite{manohar2018semi}, when model-based adaptation approaches such as LHUC \cite{Swietojanski2014LearningHU,Swietojanski2016LearningHU} are performed in an unsupervised mode. In this case, a multi-pass decoding framework is often required due to the lack of reference transcriptions of adaptation data. An initial recognition pass produces the initial speech transcription to serve as the supervision for the subsequent SD parameter estimation and re-decoding of speech in a second pass. The resulting sensitivity to the error rate of supervision hypotheses is particularly challenging with E2E systems that directly learn the surface word or token sequence labels in the adaptation supervision \cite{sivaraman2022unsupervised}, as opposed to the use of latent phonetic targets in hybrid ASR system adaptation. 

In order to address these issues, compact SD parameter representations in the forms of LHN \cite{Gemello2007LinearHT}, LHUC \cite{Swietojanski2014LearningHU,Swietojanski2016LearningHU}, hidden unit bias vectors (HUB) \cite{xue2014fast} and PAct \cite{Zhang2016DNNSA} are used in this paper to facilitate both speaker adaptive training and test-time unsupervised speaker adaptation of ESPnet Conformer models. The sensitivity to supervision quality is reduced using a more “trustworthy” subset of speaker-specific adaptation data obtained by confidence score-based selection. In Conformer models, both the attention-based encoder-decoder model architecture and auto-regressive decoder components utilize the full historical context. This makes the use of conventional lattice-based recognition hypotheses representation for word posterior and confidence score estimation non-trivial \cite{evermann2000posterior,Mangu2000FindingCI}. To this end, two types of lightweight confidence score estimation modules (CEMs) are proposed to smooth the over-confident Conformer decoder output probabilities before serving as confidence scores. The data sparsity issue, which is exacerbated by the use of confidence score-based speaker-level data selection, is addressed by replacing SD parameters with latent variable posterior distributions to model the parameter uncertainty using Bayesian learning \cite{mackay1992practical,graves2011practical,Xie2021BayesianLF}. 

Experiments are conducted on two tasks: a) the 300-hour Switchboard telephone corpus \cite{godfrey1992switchboard}; and b) the 233-hour AMI meeting speech transcription dataset \cite{hain2006ami}. Experimental results suggest that the proposed confidence score-based selection of speaker-level subset data of varying quantities produced LHUC-SAT adaptation performance comparable to the oracle, ground truth word error rate (WER) based selection. The proposed confidence score-based adaptation scheme consistently outperformed the baseline SI model and conventional non-Bayesian, point estimate-based adaptation using LHUC, HUB, or PAct. Similar consistent performance improvements were retained after external Transformer and LSTM language model rescoring. In particular, on the 300-hour Switchboard corpus, statistically significant\footnote{Matched pairs sentence-segment word error (MAPSSWE) based statistical significance test \cite{Gillick1989SomeSI} was performed at a significance level $\alpha$=0.05.} WER reductions of 1.0\%, 1.3\%, and 1.4\% absolute (9.5\%, 10.9\%, and 11.3\% relative) were obtained over the baseline SI Conformer model on the NIST Hub5’00, RT02, and RT03 evaluation sets respectively. Similar WER reductions of 2.7\% and 3.3\% absolute (8.9\% and 10.2\% relative) were also obtained on the AMI development and evaluation sets recorded under individual headset microphone (ihm), single distant microphone (sdm) and multiple distant microphones (mdm) conditions respectively. 

The main contributions of this paper are summarized below: 

1) To the best of our knowledge, this paper is the first work to investigate data efficient, compact model-based speaker adaptation approaches for E2E Conformer ASR systems. In contrast, prior research works on structured transform-based speaker adaptation were limited to the conventional hybrid DNN-HMM based ASR systems \cite{Neto1995SpeakeradaptationFH,Li2010ComparisonOD,Gemello2007LinearHT,Swietojanski2014LearningHU,Swietojanski2016LearningHU,xie2019blhuc,Xie2021BayesianLF,Zhang2016DNNSA,Zhao2016LowrankPD,xue2014fast,gales2000cluster,wu2015multi,tan2015cluster,huang2020acoustic}. In contrast, existing speaker adaptation methods designed for E2E ASR systems largely focused on using auxiliary speaker adaptive features \cite{tuske2021limit,zeineldeen2022improving,baskar2022speaker,delcroix2018auxiliary,sari2020unsupervised,fan2019speaker,zhao2020speech}, or directly adapting the whole or certain components of the SI models \cite{ochiai2018speaker,sim2021robust,huang2020rapid,huang2021rapid}. The speaker-level data sparsity issue encountered in these data-intensive model-based adaptation approaches remains largely unaddressed. Prior researches also investigated the use of auxiliary speaker embedding-based approaches for E2E systems \cite{tuske2021limit,zeineldeen2022improving,baskar2022speaker,delcroix2018auxiliary}. However, these were found to produce inconsistent results in speaker adaptation, e.g., for the IBM Conformer systems reported on the Switchboard task \cite{tuske2021limit}.  

2) This paper presents the first work on using confidence score-based unsupervised Conformer model speaker adaptation. Two forms of CEMs operating at utterance or token level, are designed to smooth the over-confident Conformer decoder output probabilities before serving as confidence scores to select the more ”trustworthy” subset of speaker-level adaptation data. Detailed implementation issues over model configuration, and choice of input features when constructing the CEMs are also investigated in a series of ablation studies. In contrast, similar E2E ASR systems’ decoder output probabilities smoothing approaches were only studied for confidence score estimation \cite{kumar2020utterance,Li2021ConfidenceEF,Liu2021UtteranceLevelNC,li2022improving}, but not for speaker adaptation of Conformer ASR systems as considered in this paper.

3) To the best of our knowledge, the research presented in this paper also presents the first investigation of full Bayesian learning powered speaker adaptation of E2E Conformer ASR systems to address the speaker-level data sparsity issue. The SD parameter uncertainty is explicitly modelled and accounted for using variational inference. In contrast, previous studies primarily used KL divergence regularization \cite{li2018speaker,meng2019speaker,weninger2019listen} and target-speaker data augmentation \cite{sim2019personalization,huang2020rapid} (by text-to-speech synthesis) to solve similar data sparsity during E2E system speaker adaptation and personalization.

4) The advantages of the proposed confidence score-based Conformer adaptation approaches are consistently demonstrated across both the Switchboard and AMI corpora over the comparable baseline non-Bayesian, point estimate-based model adaptation using a variety of SD parametric representations including LHUC, HUB, or PAct.  Their efficacy and competitiveness are further shown when being compared with state-of-the-art performance obtained on the 300-hour Switchboard task using the most recent hybrid and E2E systems reported in the literature. In addition, their generic nature and the accompanying implementation details presented in this paper allow their further application to other popular E2E ASR architectures, for example, Conformer transducers \cite{huang2021rapid}, and offer further insights into more general forms of parameter and data efficient Transformer model adaptation beyond speech processing tasks, for example, the use of task domain dependent modules when fine-tuning pre-trained language models in natural language processing \cite{houlsby2019parameter}.

The rest of this paper is organized as follows. Section \ref{sec:2} reviews the Conformer ASR systems. Section \ref{sec:3} introduces various forms of structured hidden layer transform-based speaker adaptation and their application to the Conformer model. The confidence score-based speaker adaptation scheme and Bayesian estimation are proposed in Section \ref{sec:4}. Section \ref{sec:5} discusses a set of key implementation details. Section \ref{sec:6} presents the experimental results and analysis. Finally, Section \ref{sec:7} draws the conclusions and discusses future research work. 

\section{Conformer E2E ASR Systems}\label{sec:2}
\begin{figure*}[htbp]
    \centering
    \includegraphics[scale=0.45]{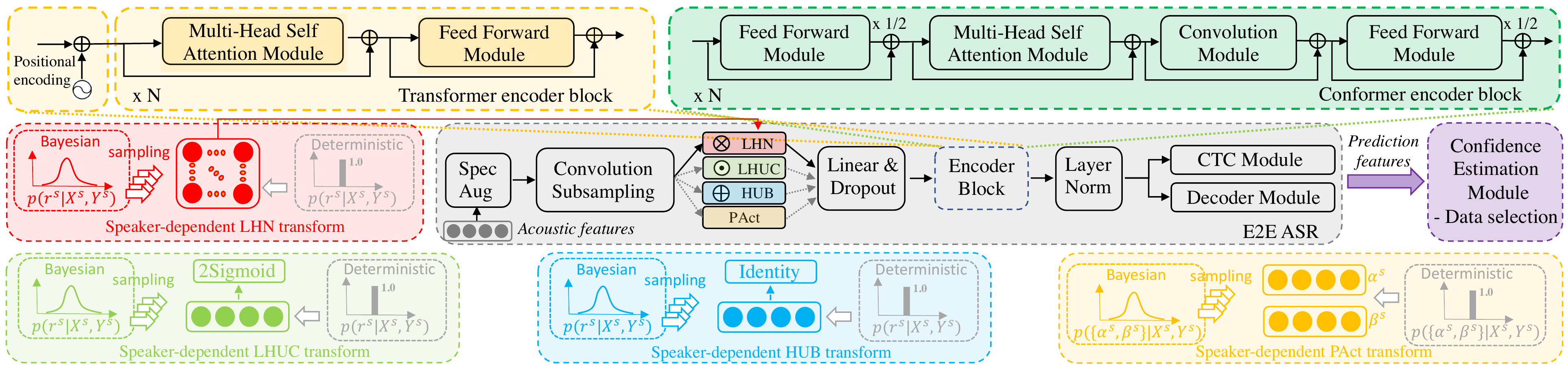}
    \setlength{\abovecaptionskip}{0pt plus 1pt minus 3pt}
    \caption{Examples of Transformer and Conformer E2E models are shown in the grey box (middle centre). Their respective Transformer and Conformer encoders are shown in the yellow (top left) and green coloured boxes (top right). The SD transformations of LHN, LHUC, HUB, and PAct are shown in the red coloured box (middle left), the light green coloured box (bottom left corner), the blue coloured box (bottom centre), and the light yellow coloured box (bottom right corner) respectively, where Bayesian estimation or fixed value, deterministic estimation are  shown on the left and right inside each box.}
    \label{fig:figure1_arch}
    \vspace{-0.5cm}
\end{figure*}

The Conformer \cite{gulati2020conformer,guo2021recent} comprises a Conformer encoder and a Transformer decoder. The Conformer encoder has one convolutional subsampling module, and a linear layer with dropout operation followed by stacked encoder blocks. The internal components of each Conformer encoder block include: a position-wise feed-forward network (FFN) module, a multi-head self-attention (MHSA) module, a convolution (CONV) module, and a final position-wise FFN module at the end. Among these, the CONV module further comprises the following layers: a 1-D point-wise convolution layer, a gated linear units (GLU) activation, a second 1-D point-wise convolution layer, a 1-D depth-wise convolution layer, a Swish activation, and a final 1-D point-wise convolution layer. All encoder blocks additionally undergo layer normalization and residual connections. Fig.~\ref{fig:figure1_arch} shows an example of Conformer E2E ASR system, where the general model architecture is in the grey colored box at the center, and the encoder block compositions are in the green colored boxes at the top right. The comparable Transformer encoder block is in the top left.

The following multitask criterion interpolation between the CTC and attention error costs \cite{watanabe2017hybrid} is utilized in training for both the Transformer and Conformer models.
\begin{equation}
{\cal{L}}=(1-\lambda) {{\cal{L}}_{att}} + \lambda{{\cal{L}}_{ctc}},
\label{enq:loss_combine}
\end{equation}
where $\lambda \in [0, 1]$ is a tunable hyper-parameter and empirically set as 0.2 for training and 0.3 for recognition in this paper.

\vspace{-0.2cm}
\section{Conformer Speaker Adaptation}\label{sec:3}
In this section, four structured transform-based speaker adaptation methods based on LHN, LHUC, HUB and PAct, along with their further integration with speaker adaptive training (SAT) are introduced for Conformer ASR systems. 

\subsection{Linear Hidden Network}
In linear hidden network (LHN) based Conformer speaker adaptation, SD affine transforms are inserted into the hidden layers of the Conformer encoder module, akin to the adaptation of hybrid DNN ASR systems using similar approaches \cite{Gemello2007LinearHT}. The LHN transform in the $l$-th hidden layer for the speaker $s$ is parameterized with a linear matrix ${\bf A}^{l,s}$ and a bias ${\bm b}^{l,s}$. The $l$-th adapted hidden layer output can be expressed by
\begin{equation}
{\bm{h}}^{l,s}={{\bf A}^{l,s}}{\bm{h}}^{l} + {{\bm b}^{l,s}},
\end{equation}
where ${\bm{h}}^{l}$ is the hidden vector after a non-linear activation function in the $l$-th layer. The linear matrix ${\bf A}^{l,s}$ and bias ${\bm b}^{l,s}$ are initialized as an identity matrix and zeros vector when performing adaptation. An example LHN based Conformer speaker adaptation is shown in the Fig.~\ref{fig:figure1_arch} (red coloured box, middle left). However, directly using a large, full matrix SD linear transformation containing millions of parameters is not only prone to over-fitting given the limited adaptation data, but also leads to computational efficiency issues. To this end, three forms of more compact SD transformations are introduced in the following subsections. 

\subsection{Learning Hidden Unit Contributions}
The key idea of learning hidden unit contributions (LHUC) adaptation \cite{Swietojanski2014LearningHU,Swietojanski2016LearningHU} is to restrict the linear SD transformation ${\bf A}^{l,s}$ as a diagonal matrix to reduce the number of SD parameters, which is equal to use a scaling vector to modify the amplitudes of activation outputs. Let ${\bm{r}}^{l,s}$ denote the SD parameters for speaker $s$ in the $l$-th hidden layer. The LHUC adapted hidden layer output can be given as 
\begin{align}
\setlength{\abovedisplayskip}{0pt plus 1pt minus 2pt}
\setlength{\belowdisplayskip}{0pt plus 1pt minus 2pt}
{\bm{h}}^{l,s}={diag({\bf A}^{l,s}}){\bm{h}}^{l}=\xi(\bm{r}^{l,s})\odot {\bm{h}}^{l},
\end{align}
where $\odot$ denotes the Hadamard product operation, and $\xi(\bm{r}^{l,s})$ is the scaling vector parameterized by ${\bm{r}}^{l,s}$. In this paper, $\xi(\cdot)$ is the element-wise 2Sigmoid$(\cdot)$ function with range $(0, 2)$ for LHUC adaptation. The SD parameters ${\bm{r}}^{l,s}$ are initialized as zeros vector during speaker adaptation. An example LHUC based Conformer speaker adaptation is shown in the Fig. \ref{fig:figure1_arch} (green coloured box, bottom left corner).

\subsection{Hidden Unit Bias}
Alternatively restricting the SD parameters ${\bm{r}}^{l,s}$ to be a bias vector that is added to the hidden layer output leads to the hidden unit bias (HUB) adaptation, akin to speaker codes \cite{xue2014fast}. Let $\xi(\bm{r}^{l,s})$ denote the bias vector parameterized by SD parameters ${\bm{r}}^{l,s}$ for speaker $s$ in the $l$-th hidden layer. The HUB adapted hidden layer output is given by 
\begin{align}
\setlength{\abovedisplayskip}{0pt plus 1pt minus 2pt}
\setlength{\belowdisplayskip}{0pt plus 1pt minus 2pt}
{\bm{h}}^{l,s}=\xi(\bm{r}^{l,s}) + {\bm{h}}^{l}.
\end{align}
In this paper, the identity activation is adopted as the bias activation function $\xi(\cdot)$ for HUB adaptation. The SD parameters $\bm{r}^{l,s}$ are initialized as zeros vector. An example HUB based Conformer speaker adaptation is shown in the Fig.~\ref{fig:figure1_arch} (blue coloured box, bottom center).

\subsection{Parameterized Activation Function}
The last speaker adaptation method used in this paper modifies the hidden activation function inputs instead of their output vectors. Considering the parameterized ReLU (PReLU) activation function proposed in \cite{Zhang2016DNNSA} for hybrid DNN systems, the speaker adapted $l$-th Conformer hidden layer PReLU activation outputs are computed in this paper as 
\begin{equation}
{{h}}^{l,s}_{n}=\begin{cases}
{\alpha}^{l,s}_{n}{z}^{l}_{n},\quad &{z}^{l}_{n}\geq 0; \\
{\beta}^{l,s}_{n}{z}^{l}_{n},\quad &{z}^{l}_{n}<0,
\end{cases} 
\end{equation}
where ${z}^{l}_{n}$ is the the $n$-th element of activation inputs $\bm{z}^{l}$, and ${\alpha}^{l,s}_n$ and ${\beta}^{l,s}_n$ are the $n$-th elements of the SD parameters ${\bm{r}}^{l,s}=\{\bm{\alpha}^{l,s},\bm{\beta}^{l,s}\}$ that correspond to slopes for positive and negative elements ${z}^{l}_{n}$ respectively. The SD parameters are initialized as $\bm{\alpha}^{l,s}={\bm{1}}$ and $\bm{\beta}^{l,s}=\bm{0}$ during adaptation, which can be viewed as the case of the standard ReLU activation. An example PAct based Conformer speaker adaptation is shown in the Fig.~\ref{fig:figure1_arch} (yellow coloured box, bottom right corner).

\subsection{Estimation of SD Parameters}
Let ${\cal{D}}^{s}\!\!=\!\!\{{\bm{X}}^{s},{\bm{Y}}^{s}\}$ denote the adaptation data set for speaker $s$, where ${\bm{X}}^{s}$ and ${\bm{Y}}^{s}$ stand for the acoustic features and the corresponding supervision token sequence, respectively. Without loss of generality and for simplicity, the Conformer SI parameters are omitted in the context of test-time adaptation. The SD parameters ${\bm{r}}^s$ can be estimated by minimizing the loss in Eqn. (\ref{enq:loss_combine}), which is given by 
\begin{align}
\setlength{\abovedisplayskip}{0pt plus 1pt minus 2pt}
\setlength{\belowdisplayskip}{0pt plus 1pt minus 2pt}
\!\!\hat{{\bm{r}}}^s&\!=\!\mathop{\arg\min}\limits_{{\bm{r}}^s}\{-\lambda_1\!\log p_{a}({\bm Y}^{s}|{\bm X}^s,\!{\bm r}^s)\!-\!\lambda\!\log p_{c}({\bm Y}^{s}|{\bm X}^s\!,\!{\bm r}^s)\},\!
\end{align}
where $\lambda_1=1-\lambda$, $p_{a}$ and $p_{c}$ are the attention and CTC based sequence probabilities, respectively. During the unsupervised test-time adaptation, the supervision ${\bm{Y}}^{s}$ of unseen test speaker $s$ needs to be generated by initially decoding the corresponding speaker-specific data using a baseline SI model, before serving as the target token labels in the subsequent adaptation stage. As a result, the estimation of SD parameters is sensitive to the underlying supervision error rate. 

During SAT training, a set of SD parameters associated with the training speakers are used to be jointly optimized with the remaining SI parameters ${{\bm{\theta}}}_{SAT}^{SI}$ based on the loss function in Eqn.~(\ref{enq:loss_combine}). This can be re-expressed as
\begin{align}
\setlength{\abovedisplayskip}{0pt plus 1pt minus 2pt}
\setlength{\belowdisplayskip}{0pt plus 1pt minus 2pt}
\{\hat{\bm{\theta}}^{SD}, \hat{{\bm{\theta}}}_{SAT}^{SI}\} = \mathop{\arg\min}_{{\bm{\theta}}^{SD}, {{\bm{\theta}}}_{SAT}^{SI}} \sum_{s\in{\cal S}} {\cal L}({\cal D}^{s}; {\bm{\theta}}^{SD}, {{\bm{\theta}}}_{SAT}^{SI}), 
\end{align}
where $\cal S$ is the set of training speakers, and $\bm{\theta}^{SD}=\{{\bm r}^{s}\}_{s\in{\cal S}}$ is the SD parameter set. This form of joint training produces a more speaker normalized and canonical SAT model that can be better tailored to unseen speakers during test-time adaptation. 

\section{Confidence Score Based Speaker Adaptation}\label{sec:4}
Confidence score-based Conformer adaptation is presented in this section. Bayesian learning-based speaker adaptation approaches are also proposed to address the increase of data sparsity as the result of confidence score-based data selection. 
\subsection{Confidence Score Based Data Selection} \label{sec41:cem}
To reduce the performance sensitivity to supervision error rate in Conformer speaker adaptation, one solution explored in this research is to use confidence score-based selection of a more “trustworthy”, accurate and less erroneous subset of speaker-specific adaptation data. Confidence scores provide a measure of the reliability of speech recognition system outputs. They have been widely adopted in speaker adaptation \cite{Anastasakos1998TheUO,Wallhoff2000FramediscriminativeAC,gollan2008confidence,Liu2013LanguageMC}, unsupervised training \cite{Zavaliagkos1998UsingUT,Kemp1999UnsupervisedTO,Yu2010UnsupervisedTA}, and semi-supervised learning \cite{srinivasamurthy2017semi} of traditional HMM-based ASR systems. 

The efficacy of confidence score-based data selection crucially depends on its correlation with the WER. Various forms of confidence measures computed at frame \cite{moreau1999use}, phoneme \cite{bartkova1997usefulness}, token or word \cite{cox1996confidence} and utterance \cite{uhrik1997confidence} levels have been investigated for speech recognition. Conventional HMM-based ASR systems often use lattice-based recognition outputs for word or token posterior probability and confidence score estimation \cite{kemp1997estimating,evermann2000posterior,wessel2001confidence}. More advanced confidence score estimation approaches that incorporate richer lattice derived features via separately designed confidence score estimators, for example, conditional random fields and recurrent neural networks, have also been studied \cite{Seigel2011CombiningIS,kalgaonkar2015estimating,kastanos2020confidence}.

However, in many current E2E ASR systems that adopt the AED model architecture and auto-regressive decoder design, including the Conformer models considered in this paper, the complete history context is normally required during inference. This presents a challenge to implement effective beam search in lattice-based decoding, unless approximations, for example, based on truncated histories are used \cite{Prabhavalkar2021LessIM}. An alternative solution is to directly interpret the Conformer decoder output probabilities or their combination with CTC probabilities as confidence scores. However, these output probabilities have been proven to be over-confident \cite{Li2021ConfidenceEF,Liu2021UtteranceLevelNC} and cannot be directly used as reliable indicators of recognition accuracy. To address this issue, two types of neural network-based CEMs that produce either utterance or token-level confidence scores are designed in this paper to smooth the over-confident Conformer decoder output probabilities, before these are used to select more accurate, less erroneous subsets of speaker-level data during test-time unsupervised adaptation. An example CEM is shown in Fig.~\ref{fig:figure2_cem}, there is one correct recognition hypothesis and one wrong recognition hypothesis. The original decoder output probabilities assigned to the correct (red rectangle in the top right) and the wrong (deep green rectangle in the top right) recognition hypotheses are 0.53 ("under-confident") and 0.93 ("over-confident"), respectively. After using the CEM to smooth these output probabilities, the probabilities of the correct (red rectangle in the bottom right) and the wrong (deep green rectangle in the bottom right) recognition hypotheses are boosted to 0.81 and reduced to 0.62 respectively, which can be considered as the more reliable confidence measures than the original decoder output probabilities. 

\subsubsection{Utterance-level CEM} 
\begin{figure}[htbp]
    \centering
    \setlength{\abovecaptionskip}{0pt plus 1pt minus 3pt}
    \includegraphics[scale=0.5]{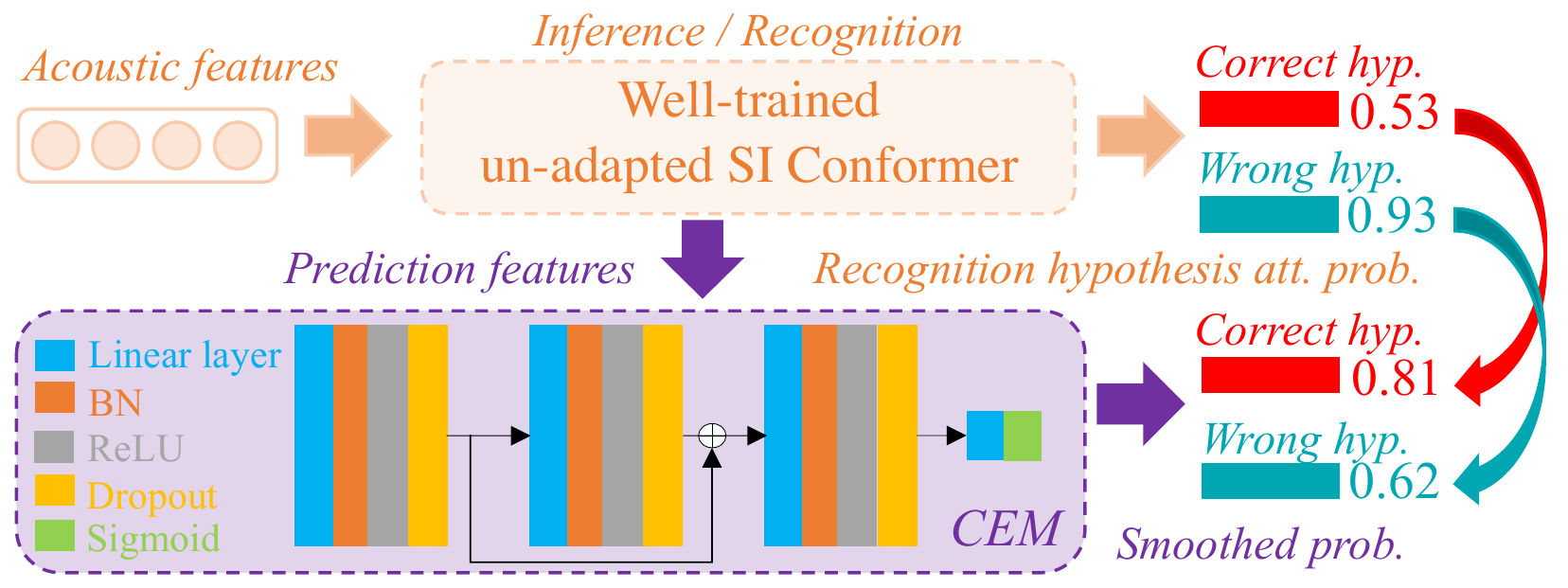}
    \caption{An example CEM architecture is shown in the purple box. The prediction features are extracted from different internal locations of a Conformer model (light orange box), as discussed in Section \ref{sec:51feature_cem}.}
    \label{fig:figure2_cem}
\vspace{-0.5cm}
\end{figure}

As shown in Fig.~\ref{fig:figure2_cem}, the form of utterance-level CEM is a lightweight binary classification model, which contains a simple yet effective 3-layer residual feed-forward network with a hidden dimension of 64 and a Sigmoid output layer. Batch normalization, ReLU activation, and dropout neural operations are applied to each hidden layer in sequence. Skip connection is also utilized between the outputs of the first and the second hidden layer. This CEM is then connected with ASR systems such as a SI Conformer model. A set of prediction features derived from different internal components of the Conformer model during the recognition process is fed into the CEM that subsequently predicts a smoothed confidence score $c_{k}$ for the current utterance $k$. It is noted that the prediction features play an important role in the design of robust CEMs, more details of prediction feature selection will be provided in Section \ref{sec:51feature_cem}. To produce the target utterance-level confidence score labels for CEM training, recognition hypotheses are produced by performing beam search decoding of the training data using an un-adapted SI Conformer model. The binary target label $l_k$ of utterance $k$ is then assigned either a value of one if its recognition hypothesis exactly matches the corresponding ground-truth reference, or zero otherwise. The following cross-entropy error cost computed between the confidence score $\bm c$ and the target label $\bm {l}$ is used to optimize the CEM. 
\begin{align}
\setlength{\abovedisplayskip}{0pt plus 1pt minus 2pt}
\setlength{\belowdisplayskip}{0pt plus 1pt minus 2pt}
{\cal L}_{c} = \sum\limits_{k\in {\cal B}}{(\eta l_k \log(c_k) + (1-\eta)(1-l_k)\log(1-c_k))}, \label{loss_cem}
\end{align}
where ${\cal B}$ is the set of training samples, and $\eta\in [0, 1]$ is a weighting factor empirically set as $0.3$ in this paper. 

\subsubsection{Token-level CEM} 
The token-level CEM shares the same overall architecture with the utterance-level CEM, while the Sigmoid output layer produces a confidence score indicating the probability of the current token being correct based on the more fine-grained token-level features. As the speaker-level data selection operates at the utterance level, utterance-level confidence scores can be obtained by mean averaging the token-level confidence measures within each utterance. 

The target labels for token-level CEM training are determined as follows. The edit distance is first used to produce the alignment between the recognition hypothesis and the ground-truth reference. Based on such alignment, the correct token is assigned a label of one, while the substituted or inserted token is labelled with zero. For example, if the ground-truth reference token sequence is "C U H K" and the corresponding recognition hypothesis is "C U L K", then the label sequence is ${\bm l}=[1, 1, 0, 1]$. The resulting binary labelled CEM training data is down sampled to balance the numbers of tokens with one and zero confidence scores\footnote{For example, the 300-hour Switchboard training set contains a total of 1,267,693 tokens, among which 1,163,601 and 104,092 are labelled as one and zero respectively. The respective number of one and zero labelled tokens used in CEM training are 416,368 and 104,092.}. The binary cross-entropy loss similar to Eqn.~(\ref{loss_cem}) is also used here, except that the loss is computed at a token level. 

\subsection{Bayesian Learning of Adaptation Approach}
Despite confidence score-based data selection can alleviate the negative effect on adaptation from supervision errors, it further reduces the amount of already limited speaker-level adaptation available data, for example, a few minutes of Switchboard conversational speech per speaker, and exacerbates the data sparsity issue. This leads to SD parameter uncertainty during the standard point estimate-based model adaptation, even for the more compact LHUC, HUB and PAct adaptation of Sec. \ref{sec:3}. To address this issue, Bayesian learning \cite{mackay1992practical,graves2011practical,Xie2021BayesianLF} by using a posterior distribution to model the parameter uncertainty is adopted in this paper. The Bayesian predictive distribution for the test utterance ${\tilde{\bm X}}^s$ is obtained by integrating over SD parameter values as 
\begin{equation}
p(\tilde{\bm Y}^s|\tilde{\bm X}^s,{\cal D}^s)=\int{p(\tilde{\bm{Y}}^{s}|\tilde{\bm{X}}^{s},{\bm r}^s)p({\bm r}^s|{\cal D}^s)d{\bm r}^s},
\label{eq:posterior}
\end{equation}
where ${\tilde{\bm Y}}^s$ is the predicted token sequence and $p({\bm r}^s|{\cal D}^s)$ is the SD parameters posterior distribution learned from adaptation data of speaker $s$. The hybrid attention plus CTC loss marginalisation over the adaptation data ${\cal D}^s$ is optimized to infer the posterior distribution over the SD latent variable ${\bm r}^s$. 
\begin{align}
\setlength{\abovedisplayskip}{0pt plus 1pt minus 2pt}
\setlength{\belowdisplayskip}{0pt plus 1pt minus 2pt}
&\!\!{\cal L}({\cal D}^{s}) = -\lambda_1 \log p_{a}({\bm Y}^{s}|{\bm X}^s) - \lambda \log p_{c}({\bm Y}^{s}|{\bm X}^s) \nonumber\\
&\!\!\!\!=\!\!\int \!{\left\{-\lambda_1\log p_{a}({\bm Y}^{s},{\bm r}^s|{\bm X}^s)\! - \!\lambda \log p_{c}({\bm Y}^{s},{\bm r}^s|{\bm X}^s)\right\}\! d{\bm r}^s}. \label{evde_int}
\end{align}
The commonly used variational inference is applied to learn the SD parameter posterior distribution in the above integration. A variational distribution $q(\bm r^s)$ is introduced to approximate the posterior distribution $p({\bm r}^s|{\cal D}^s)$ by minimising the following upper bound of the integration in Eqn.~(\ref{evde_int}) derived by Jensen’s inequality, which is given as 
\begin{align}
\setlength{\abovedisplayskip}{0pt plus 1pt minus 2pt}
\setlength{\belowdisplayskip}{0pt plus 1pt minus 2pt}
&{\cal L}({\cal D}^{s})\! \leq\! \!\!\int{\!\!\!q({\bm r}^s)\!\left\{\!-\!\lambda_1\!\log p_{a}\!({\bm Y}^{s}\!|\!{\bm X}^s\!,\!{\bm r}^s)\!\!-\!\!\lambda\!\log p_{c}({\bm Y}^{s}\!|\!{\bm X}^s\!,\!{\bm r}^s)\!\right\}\!d{\bm r}^s} \nonumber \\
& + {\text {KL}}(q({\bm r}^s)||p({\bm r}^s)) \triangleq {\cal L}_1 + {\cal L}_2, \label{upebound}
\end{align}
where $p({\bm r}^s)$ denotes the prior distribution of SD parameters, ${\text {KL}}(\cdot)$ is the KL divergence. To ensure that the integration ${\cal L}_1$ is differentiable, the Monte Carlo sampling method with a re-parameterization trick is used to approximate the term ${\cal L}_1$, 
\begin{equation}
{\cal L}_1\approx \frac{1}{N}\!\sum\limits_{n=1}^{N}{\!-\lambda_1\!\log p_{a}({\bm Y}^{s}|{\bm X}^s\!,\!{\bm r}_n^s)\!-\!\lambda\!\log p_{c}({\bm Y}^{s}|{\bm X}^s\!,\!{\bm r}_n^s)},
\label{eq:loss1}
\end{equation} %
where ${\bm r}_n^s={\bm \mu}+{\bm \sigma}\odot {\bm \epsilon}_n$ and ${\bm \epsilon}_n$ is the $n$-th Monte Carlo sampling vector drawn from a standard normal distribution. For simplicity, assuming that both $q({\bm r}^s)={\cal N}({\bm \mu}, {\bm \sigma}^2)$ and $p({\bm r}^s)={\cal N}(\bar{\bm \mu}, \bar{\bm \sigma}^2)$ follow the normal distributions. The KL divergence ${\cal L}_2$ can be explicitly expressed as a closed form, 
\begin{equation}
\mathcal{L}_2=\frac{1}{2}\sum_{k}{\left(\frac{\sigma_k^2+(\mu_k-\bar{\mu}_{k})^2}{\bar{\sigma}_{k}^2}+2\log\frac{\bar{\sigma}_{k}}{\sigma_{k}}-1\right)},
\end{equation}
where $\{\bar \mu_{k}, \bar\sigma_{k}\}$ and $\{\mu_k,\sigma_k\}$ are the $k$-th elements of the corresponding vectors $\{\bar{\bm \mu}, \bar{\bm \sigma}\}$ and $\{{\bm \mu},{\bm \sigma}\}$, respectively. 

\section{Implementation Details} \label{sec:5}
A set of important implementation issues that affect the performance and efficiency of the proposed confidence score-based Conformer speaker adaptation approaches are discussed in this section. 
\subsection{Feature Selection for CEM} \label{sec:51feature_cem}
A key issue in designing robust CEMs is to select the most informative prediction features to better distinguish the correct recognition hypotheses from erroneous ones. Inspired by previous research on confidence score estimation for E2E ASR systems \cite{kumar2020utterance,li2022improving}, various forms of neural features extracted from different components of the Conformer ASR system for CEM training are investigated in this paper. These include: \textbf{1) Internal embedding features} that are extracted at the last layers of the encoder and decoder blocks; \textbf{2) 1-best output scores} that are produced during beam search by either the decoder or CTC module alone, or their combination, for the current 1-best token, or utterance-level hypothesis being considered; \textbf{3) N-best output scores} as an extension to 2) that can exploit additional information of modelling confusion over competing hypotheses or tokens; and \textbf{4) Feature fusion} among the above from 1) to 3). A detailed ablation study on feature selection for CEM training and confidence score estimation is presented in the following Sec. \ref{sec63:cem}. 

\subsection{The Location of SD Transformations}\label{sec51:location}
The fundamental task of SD parameters is to account for speaker-level variability exhibited among the Conformer hidden layer outputs. In common with many other deep neural network systems, the higher Conformer encoder layers that are closer to the encoder output tend to produce more denoised and perturbation invariant features than those positioned at the bottom. As a result, the speaker-level variability that manifested in the lower Conformer hidden layer outputs is expected to be larger, and needs to be accounted for using SD parameters, than that found among those at the 
higher layers. This is confirmed in a set of ablation study experiments over different Conformer encoder internal locations to introduce LHUC transforms, as presented later in Table \ref{tab2: bayes_config} of Sec. \ref{sec62:ablation}. In practice, it is observed that applying the SD LHUC transformation to the convolution subsampling module produced the best configuration speaker adaptation performance among different locations inside the encoder. 

\subsection{Choice of Prior Distributions}
When performing Bayesian estimation of SD parameters, appropriate parameter prior distributions are required to be set based on the form of the corresponding activation functions in different adaptation approaches. The prior distribution can be set to be a specific simple normal distribution. In this paper, the prior distributions for LHUC, HUB, and PAct adaptations are set to be $p({\bm r}_s)={\cal N}({\bf 0}, {\bf 1})$, $p({\bm r}_s)={\cal N}(\bf 0, \bf 0.001)$ and $\left\{p({\bm \alpha}_s), p({\bm \beta}_s)\right\}=\left\{{\cal N}(\bf 1, \bf 1),{\cal N}(\bf 0, \bf 1)\right\}$ respectively. Alternatively, an empirical distribution estimated from training data can be used as the prior distribution for Bayesian estimation. The prior distribution ${\cal N}(\bar{\bm \mu}, \bar{\bm \sigma}^2)$ can be computed over the SD transformations of all training speakers, which is given as 
\begin{align}
\bar{\bm \mu}=\frac{1}{\left|{\cal S}\right|}{\mathop{\sum_{s\in {\cal S}}{\bm r}_s}},~\bar{\bm \sigma}^2=\frac{1}{\left|{\cal S}\right|}{\mathop{\sum_{s\in {\cal S}} \{({\bm r}_s}} - \bar{\bm{\mu}}) \odot  ({\bm r}_s - \bar{\bm{\mu}})\}, \label{prior_train}
\end{align}
where $\left|{\cal S}\right|$ is the number of entities of training speaker set $\cal S$.

\subsection{Parameter Sampling} \label{sec53:sample}
Another important issue for Bayesian estimation is the computational efficiency of inference. The variational inference used in Bayesian adaptation requires the Monte Carlo sampling vectors to approximate the first term of the upper bound in Eqn.~(\ref{eq:loss1}) during adaptation and the integration in Eqn.~(\ref{eq:posterior}) during evaluation. Since the samples are drawn independently for each parameter value, the resulting inference cost during adaptation grows linearly as the number of samples. In order to ensure that the inference cost of Bayesian adaptation is comparable to that of conventional non-Bayesian, point estimate-based adaptation, only one sample is drawn in Eqn.~(\ref{eq:loss1}) to update the SD parameters during adaptation. This is supported through an ablation study of the number of parameter samples in Table \ref{tab2: bayes_config} to be presented in Sec. \ref{sec62:ablation}. During the evaluation, the integration in Eqn.~(\ref{eq:posterior}) required by Bayesian adaptation can be efficiently approximated using the variational distribution’s mean vector.
\begin{align}
\vspace{-0.2cm}
\setlength{\abovedisplayskip}{0pt plus 1pt minus 2pt}
\setlength{\belowdisplayskip}{0pt plus 1pt minus 2pt}
\int{p(\tilde{\bm{Y}}^{s}|\tilde{\bm{X}}^{s},{\bm r}^s)p({\bm r}^s|{\cal D}^s)d{\bm r}^s}\approx p(\tilde{\bm{Y}}^{s}|\tilde{\bm{X}}^{s},{\bm \mu}).
\end{align}

\section{Experiments} \label{sec:6}
In this section, the proposed confidence score-based adaptation schemes using various SD parameter representations in the forms of LHN, LHUC, HUB and PAct are investigated for test-time unsupervised adaptation for SI and SAT Conformer systems on two tasks: the widely used 300-hour Switchboard conversational telephone corpus \cite{godfrey1992switchboard} and the speed perturbed 233-hour AMI meeting speech transcription dataset \cite{hain2006ami}. For all SAT based Conformer systems, supervised adaptation to training speaker data with no data selection is used.

This section is organized as follows. The concrete experimental setups and results of Switchboard and AMI tasks are shown in Section \ref{sec6:swbd} and Section \ref{sec6:ami}, respectively. For the 300-hour Switchboard task, Section \ref{sec61:setup} introduces the Switchboard speech dataset and the detailed configurations of the baseline Transformer and Conformer ASR systems. Section \ref{sec62:ablation} presents the performance of various LHN, LHUC, HUB, and PAct adaptation approaches under different settings. These preliminary experimental findings serve to confirm the implementation details and settings discussed in Section \ref{sec:5}, which are then adopted in the following experiments. In Section \ref{sec63:cem}, the effectiveness of the proposed confidence score-based data selection is evaluated through an extensive comparison with various selection approaches using different confidence measures. Finally, in Section \ref{sec64:main}, the advantages of the proposed confidence score-based Bayesian adaptation are demonstrated by performing various forms of structured transformations before and after external language model rescoring and adding speed perturbation-based training data augmentation. The competitiveness of the proposed adaptation is further presented by comparison with the state-of-the-art performance achieved on the 300-hour Switchboard task using the hybrid and E2E ASR systems reported in the latest literature. For the 233-hour AMI task, the AMI meeting speech transcription dataset recorded under the ihm, sdm, and mdm conditions and the baseline Conformer ASR system are described in Section \ref{sec6b1:setup}. The performance of the proposed confidence score-based adaptation scheme evaluated on the AMI set is further illustrated in Section \ref{sec6b2:main} by conducting a comparable set of experiments. All of our models were trained with one thread on a single NVIDIA Ampere A40 GPU. 

\vspace{-0.2cm}
\subsection{Experiments on Switchboard Data}\label{sec6:swbd}
\subsubsection{Experimental Setup} \label{sec61:setup}
 The Switchboard-1 conversational telephone corpus (LDC97S62) \cite{godfrey1992switchboard} with approximately 300-hour audio data recorded from 4804 speakers is utilized for training. For performance evaluation, the NIST Hub5’00 (LDC2002S09, LDC2002T43), RT02 (LDC2004S11), and RT03 (LDC2007S10) test sets collected from 80, 120, and 144 speakers respectively are used. The 3.8-hour Hub5’00 speech data consists of two subsets: the CallHome (CHE) with 40 speakers and the Switchboard (SWBD) with 40 speakers. The 6.4-hour RT02 set consists of three subsets: the Switchboard 1 (SWBD1), the Switchboard 2 (SWBD2), and the Switchboard Cellular (SWBD3). Each subset contains 40 speakers. The 6.2-hour RT03 data consists of two subsets: the Fisher (FSH) and the Switchboard (SWBD). Each subset contains 72 speakers. 

The baseline Transformer and Conformer models consist of 12 encoder and 6 decoder blocks following the ESPnet recipe\footnote{github.com/espnet/espnet/blob/master/egs/swbd/asr1/run.sh}. Each encoder or decoder block is configured with 4-head attention of 256 dimensions and 2048 feed-forward hidden units. The convolutional subsampling module includes two 2D convolutional layers with a stride of size 2, each followed by a ReLU activation. 80-dim Mel-filter bank plus 3-dim pitch parameters and Byte-pair-encoding (BPE) tokens of size 2000 are utilized as input features and the decoder outputs, respectively. SpecAugment technique \cite{Park2019SpecAugmentAS} is employed in both SI and SAT model training. The initial learning rate of the Noam optimizer is 5.0. The dropout rate is set to 0.1. The training epoch is 100, and the final recognition model for evaluation is obtained by averaging the last ten epochs. The log-linearly interpolated external Transformer and Bidirectional LSTM (BiLSTM) language models (LMs) trained on the Switchboard and Fisher transcripts (LDC2004T19, LDC2005T19) using cross-utterance contexts \cite{Sun2021TransformerLM} are further used to rescore the 30-best hypotheses generated by the Conformer ASR systems.

\begin{table}[!t]
\centering
\setlength{\abovecaptionskip}{0pt plus 1pt minus 4pt}
\caption{Performance (WER\%) of Transformer and Conformer systems using different adaptation methods evaluated on the Hub5'00 test sets. "O.V." stands for "overall". "$\dagger$" denotes a statistically significant difference obtained over the corresponding baseline SI Transformer (sys. 1) or Conformer (sys. 10) systems.}
\label{tab1: trans_conf_system}
\resizebox{0.95\columnwidth}{!}{
\begin{tabular}{c|c|l|c|ccc} %
	\hline \hline
    \multirow{2}{*}{ID} & 
    \multirow{2}{*}{Model} &
    \multirow{2}{*}{\tabincell{c}{Adaptation \\ Method}} &
     \multirow{2}{*}{\tabincell{c}{\# SD Parameters \\ Per Speaker}} &
    \multicolumn{3}{c}{Hub5'00 } 
    \\   \cline{5-7}
	& & & & CHE & SWBD & O.V.   \\ \hline\hline
	1 & \multirow{9}{*}{Transformer} & - & - & $17.3$ & $9.1$ & $13.2$ \\ 
	2 & & LHN  & $26,214,400~(100\%)$ & $17.9$ & $9.5$ & $13.7$  \\
	3 & & LHUC & $5,120~(\sim0.02\%)$ & $16.8^{\dagger}$ & $9.1$ & $13.0$ \\ 
	4 & & HUB  & $5,120~(\sim0.02\%)$ & $16.9$ & $9.1$ & $13.0$  \\ 
	5 & & PAct & $10,240~(\sim0.04\%)$ & $17.0$ & $9.1$ & $13.1$  \\ 

	6 & & LHN-SAT  &$26,214,400~(100\%)$ & $18.0$ & $9.6$ & $13.8$ \\
	7 & & LHUC-SAT & $5,120~(\sim0.02\%)$ & $\textbf {16.1}^{\dagger}$ & $\textbf {8.5}^{\dagger}$ & $\textbf {12.3}^{\dagger}$ \\ 
	8 & & HUB-SAT  & $5,120~(\sim0.02\%)$ & $16.2^{\dagger}$ & $\textbf {8.5}^{\dagger}$ & $\textbf {12.3}^{\dagger}$ \\ 
	9 & & PAct-SAT & $10,240~(\sim0.04\%)$ & $16.5^{\dagger}$ & 8.9 & $12.8^{\dagger}$ \\ 

	\hline \hline
	
	10 & \multirow{9}{*}{Conformer} & - & - & $15.0$ & $7.3$ & $11.1$ \\
	11 & & LHN  & $26,214,400~(100\%)$ & $15.3$ & $7.4$ & $11.3$ \\
	12 & & LHUC & $5,120~(\sim0.02\%)$ & $14.9$ & $7.3$ & $11.1$ \\ 
	13 & & HUB  & $5,120~(\sim0.02\%)$ & $14.9$ & $7.3$ & $11.1$ \\ 
	14 & & PAct & $10,240~(\sim0.04\%)$ & $15.0$ & $7.2$ & $11.1$ \\ 

	15 & & LHN-SAT  & $26,214,400~(100\%)$ & $15.2$ & $7.5$ & $11.4$ \\
	16 & & LHUC-SAT & $5,120~(\sim0.02\%)$ & $\textbf {13.8}^{\dagger}$ & $\textbf {7.1}$ & $\textbf {10.5}^{\dagger}$ \\ 
	17 & & HUB-SAT  & $5,120~(\sim0.02\%)$ & ${14.2}^{\dagger}$ & $\textbf {7.1}$ & ${10.7}^{\dagger}$ \\ 
	18 & & PAct-SAT & $10,240~(\sim0.04\%)$ & $14.7$ & $7.2$ & $ 11.0^{\dagger}$ \\

	\hline \hline

\end{tabular} 
}
\vspace{-0.5cm}
\end{table}

\begin{table}[!t]
\centering
\setlength{\abovecaptionskip}{0pt plus 1pt minus 4pt}
\caption{Performance (WER\%) of LHUC-SAT Conformer systems on the Hub5'00 test set with different configurations including: locations to apply LHUC transforms during SAT and test-time adaptation (col. 3-4); the SD parameter prior distribution in Bayesian adaptation (col. 5); and the number of parameter samples in variational inference (col. 6). "Conv2d", "Enc. (X)", and "Enc. (X-Y)" denote the convolution subsampling module, "x-th" encoder block layer, and the encoder block layer from "x-th" to "y-th", respectively.}
\label{tab2: bayes_config}
\resizebox{0.95\columnwidth}{!}{
\begin{tabular}{c|c|l|l|c|c|ccc} %
	\hline \hline
    \multirow{2}{*}{ID} & 
    \multirow{2}{*}{\tabincell{c}{Adapt.\\ Param.}} &
    \multicolumn{2}{c|}{Location} &
    \multirow{2}{*}{Prior} & 
    \multirow{2}{*}{\tabincell{c}{\#Sa-\\mples} } & 
    \multicolumn{3}{c}{Hub5'00} 
    \\  \cline{3-4} \cline{7-9}
	& & SAT & Test-time & & & CHE & SWBD & O.V.  \\ \hline\hline

	1 &- & - & - & - & - & 15.0 & 7.3 & 11.1  \\ \hline 
	2 & \multirow{18}{*}{\tabincell{c}{Deter- \\ ministic}} & \multirow{4}{*}{Conv2d} & Conv2d & \multirow{4}{*}{-} & \multirow{4}{*}{-} & \textbf{13.8} & 7.1 & \textbf{10.5} \\ 
	3 &  &  & Enc. (1) & &  & 14.9 & 7.4 & 11.2 \\ 
	4 &  &  & Enc. (6) & &  & 15.5 & 7.5 & 11.5 \\ 
    5 &  &  & Enc. (12) & &  & 16.5 & 7.8 & 12.2 \\ \cline{3-9}
	6 &  & \multirow{4}{*}{Enc. (1)} & Conv2d & \multirow{4}{*}{-} & \multirow{4}{*}{-} & 15.1 & 7.4 & 11.3 \\ 
	7 &  &  & Enc. (1) & &  & 14.3 & 7.1 & 10.7 \\ 
	8 &  &  & Enc. (6) & &  & 16.1 & 7.7 & 12.0 \\ 
    9 &  &  & Enc. (12) & &  & 16.6 & 8.1 & 12.4 \\ \cline{3-9}
	10 &  & \multirow{4}{*}{Enc. (6)} & Conv2d & \multirow{4}{*}{-} &\multirow{4}{*}{-} & 16.7 & 7.6 & 12.2 \\ 
	11 &  &  & Enc. (1) & &  & 15.1 & 7.2 & 11.2 \\ 
	12 &  &  & Enc. (6) & &  & 14.5 & 7.2 & 10.9 \\ 
    13 &  &  & Enc. (12) & & & 16.7 & 8.2 & 12.5 \\ \cline{3-9}
	14 &  & \multirow{4}{*}{Enc. (12)} & Conv2d & \multirow{4}{*}{-} & \multirow{4}{*}{-} & 16.8 & 7.9 & 12.4 \\ 
	15 &  &  & Enc. (1) & &  & 15.6 & 7.8 & 11.7 \\ 
	16 &  &  & Enc. (6) & &  & 15.4 & 7.6 & 11.5 \\ 
    17 &  &  & Enc. (12) & &  & 14.7 & 7.3 & 11.0 \\ \cline{3-9}
	18 &  & \multicolumn{2}{c|}{Conv2d \& Enc. (1)} & \multirow{3}{*}{-} & \multirow{3}{*}{-} & 13.9 & \textbf{7.0} & \textbf{10.5} \\
	19 &  & \multicolumn{2}{c|}{Conv2d \& Enc. (1-6)} &  &  & 14.1 & 7.1 & 10.6 \\
	20 &  & \multicolumn{2}{c|}{Conv2d \& Enc. (1-12)} &  &  & 14.1 & 7.2 & 10.7 
	\\ \hline
	21 & \multirow{6}{*}{Bayes} &
	\multicolumn{2}{c|}{\multirow{6}{*}{Conv2d}}
	& Empiri. & 1 & 13.8 & 7.0 & 10.4  \\ \cline{5-5}
	22 &  & \multicolumn{2}{c|}{} & \multirow{5}{*}{${\cal{N}}(\bf{0},\bf{1})$} & 1 & \textbf{13.7} & \textbf{6.9} & \textbf{10.3} \\ 
	23 & & \multicolumn{2}{c|}{} &  & 2 & 13.8 & 6.9 & 10.3   \\ 
	24 & & \multicolumn{2}{c|}{} &  & 3 & 13.8 & 7.0 & 10.4 \\ 
	25 & & \multicolumn{2}{c|}{} &  & 4 & \textbf{13.7} & 7.0 & \textbf{10.3}  \\ 
	26 & & \multicolumn{2}{c|}{} &  & 5 & 13.8 & 6.9 & 10.3 \\ 

	\hline \hline
\end{tabular} }
\vspace{-0.6cm}
\end{table}

\subsubsection{Performance of Baseline Adaptation} \label{sec62:ablation}
The performance of non-Bayesian, point estimate-based adaptation approaches using LHN, LHUC, HUB and PAct with or without incorporating SAT for both Transformer and Conformer systems evaluated on the Hub5’00 set is shown in Table \ref{tab1: trans_conf_system}. All speaker data was used in adaptation. The SD transformations were applied at the convolution subsampling module. For both the Transformer and Conformer systems, several trends can be observed: \textbf{a)} Test-time adaptation of non-SAT systems using LHUC, HUB or PAct produced no significant performance improvement compared to the baseline SI systems (sys. 3-5 vs. sys. 1 and sys. 12-14 vs. sys. 10). By further incorporating SAT training, the resulting LHUC-SAT, HUB-SAT, and PAct-SAT systems consistently produced statistically significant overall WER reductions of up to 0.9\% absolute over the baseline SI systems (sys. 7-9 vs. sys. 1 and sys. 16-18 vs. sys. 10). This confirms that SAT models can be better adapted to unseen speakers than SI models during test-time adaptation. \textbf{b)} The LHN adapted systems performed worse than the SI systems regardless of whether SAT was applied or not (sys. 2,6 vs. sys. 1 and sys. 11,15 vs. sys. 10). This is because LHN systems with too many SD parameters are prone to over-fitting when given limited adaptation data, as indicated in Table \ref{tab1: trans_conf_system} (col. 4). \textbf{c)} The adapted LHUC-SAT Conformer system (sys. 16) produced the lowest WERs and was selected to conduct the subsequent ablation studies on the implementation issues regarding SD transformation locations and Bayesian adaptation, as set out from Sec. \ref{sec51:location} to \ref{sec53:sample}.

Table \ref{tab2: bayes_config} shows the performance of adapted LHUC-SAT Conformer systems with different configurations, as set out in the implementation details from Sec. \ref{sec51:location} to \ref{sec53:sample}. Several findings can be found. \textbf{a)} Applying the SD transformations consistently at the same layer (sys. 2,7,12,17) during SAT and test-time adaptation achieved better performance than otherwise. Moreover, the SD transformations only need to be applied at the convolution subsampling module (sys. 2) which is close to the raw input features, as further applying SD transformations at any subsequent higher layers (sys. 18-20) produced no additional improvement. This confirms that the speaker variability found in the lower Conformer hidden layer outputs is larger than that of higher layers, as discussed in Sec. \ref{sec51:location}. \textbf{b)} The Bayesian estimation using the zero mean and unit variance based normal prior distribution (sys. 22) produced better performance than that using the empirical prior distribution (sys. 21). \textbf{c)} It is reasonable to draw only one SD parameter sample for the computation in Eqn.~(\ref{eq:loss1}) as previously suggested in Sec. \ref{sec53:sample}. No statistically significant WER difference was observed when drawing more Monte Carlo samples during adaptation (sys. 22-26). The best performing set of adaptation configurations (sys. 22) was used for the following experiments in the rest of this paper. 

\begin{figure*}[htbp]
\begin{minipage}[t]{0.335\linewidth}
\centering
\includegraphics[width=2.38in]{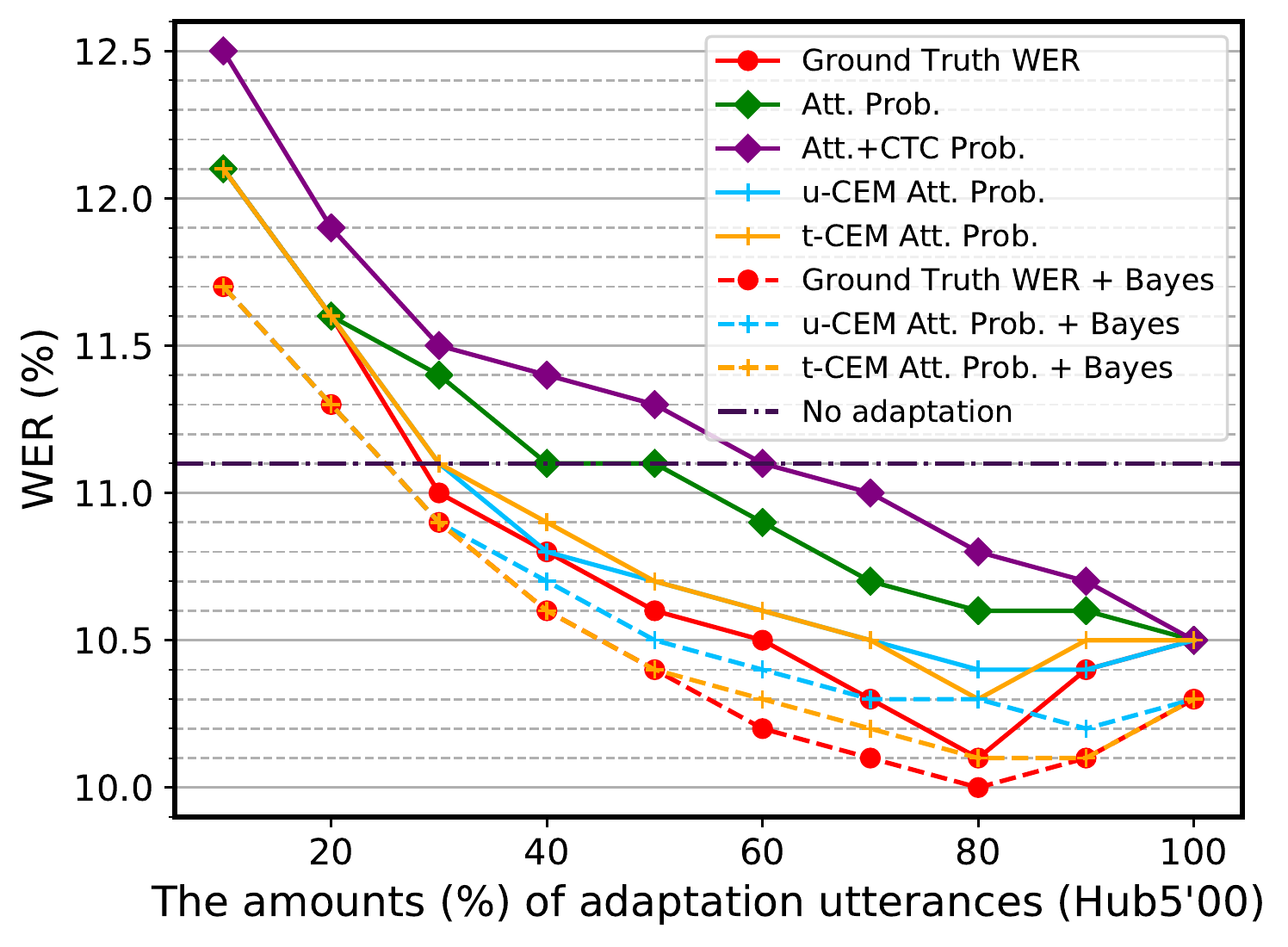}
\end{minipage}%
\begin{minipage}[t]{0.33\linewidth}
\centering
\includegraphics[width=2.35in]{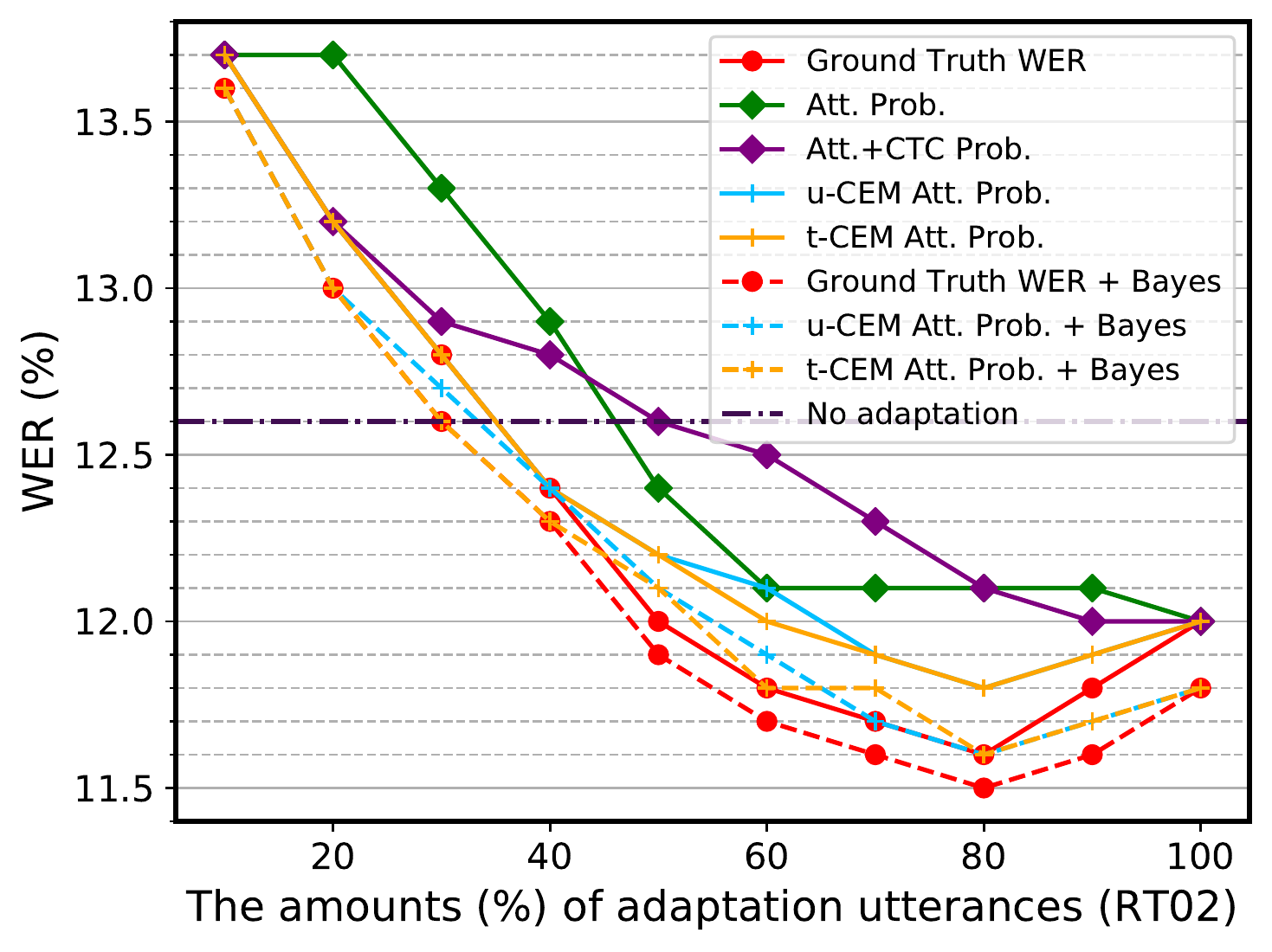}
\end{minipage}
\begin{minipage}[t]{0.30\linewidth}
\centering
\includegraphics[width=2.35in]{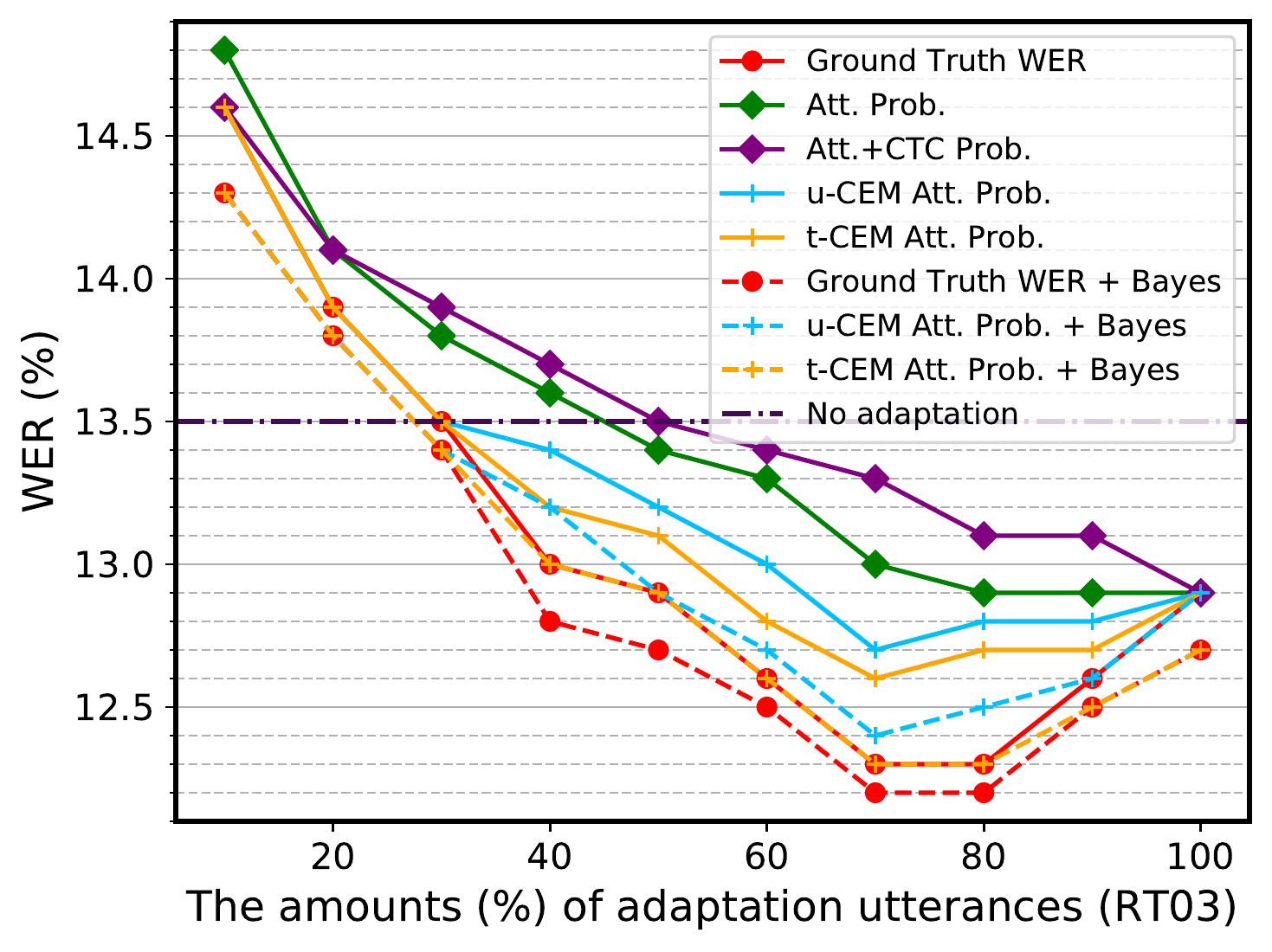}
\end{minipage}

\centering
\setlength{\abovecaptionskip}{0pt plus 1pt minus 6pt}
\caption{Performance (WER\%) of adapted LHUC-SAT Conformer systems with varying amounts of adaptation data selected using different confidence measures (top down over legends): ground truth WER ”Ground Truth WER”; original Conformer decoder attention Softmax output log-probabilities “Att. Prob.”, or their combination with CTC log-probabilities with a ratio of 7:3 “Att+CTC Prob.”; the utterance-level CEM smoothed attention probabilities “u-CEM Att. Prob.”; the token-level CEM smoothed attention probabilities “t-CEM Att. Prob.”; “+ Bayes” stands for Bayesian test-time adaptation. The horizontal coordinates stand for speaker-level selected subsets of utterances of various sizes that are ranked by their confidence measures and used for adaptation.}
\label{fig:cem}
\vspace{-0.5cm}
\end{figure*}

\begin{table}[!t]
\centering
\setlength{\abovecaptionskip}{0pt plus 1pt minus 4pt}
\caption{Performance of utterance and token level CEMs using different prediction features evaluated on the Hub5'00 data. The "Encoder embedding" and "Decoder embedding" are the hidden vectors extracted at the last encoder and decoder block layers, respectively."(Average)" means that the feature is averaged over the frames or output tokens. "10-best attention scores", "10-best CTC scores" and "10-best attention+CTC (7:3) scores" are the negative log-probabilities of 10-best recognition hypotheses obtained from the decoder module, the CTC module, and their combination with a weighting of 7:3, respectively. "Top-10 score logits" denotes the linear outputs corresponding to Top-10 decoder softmax probabilities over the entire vocabulary.}
\label{tab31:predictorfeatures}
\resizebox{\columnwidth}{!}{
\begin{tabular}{c|c|c|l|c|c} %
	\hline \hline
    \multirow{1}{*}{ID} & 
    \multirow{1}{*}{CEM} &
    \multicolumn{2}{c|}{Prediction features} & 
    \multirow{1}{*}{AUC} & 
    \multirow{1}{*}{EER} \\
    \hline \hline
    1 & \multirow{11}{*}{\tabincell{c}{Utterance \\ level}} & \multirow{2}{*}{\tabincell{c}{Internal \\ embedding}} & Encoder embedding (Average) & 0.7648 & 0.3232 \\  
    2 &  & & Decoder embedding (Average) & \textbf{0.7672} & \textbf{0.3194} \\  \cline{3-3}
    3 &  & \multirow{6}{*}{Score} &1-best attention score & 0.7902 & 0.2888 \\  
    4 &  & &5-best attention scores & 0.8524 & 0.2302 \\  
    5 &  & &10-best attention scores & \textbf{0.8538} & \textbf{0.2279} \\  
    6 &  & &20-best attention scores & 0.8501 & 0.2291 \\  
    7 &  & &10-best CTC scores & 0.8227 & 0.2634 \\  
    8 &  & &10-best attention+CTC (7:3) scores & 0.8491 & 0.2375 \\  \cline{3-3} 
    9 &  &\multirow{3}{*}{Fusion} & ID (1) + ID (5) & 0.8603 & 0.2233 \\  
    10 &  & & ID (2) + ID (5) & \textbf{0.8627} & \textbf{0.2204} \\  
    11 &  & & ID (1) + ID (2) + ID (5) & 0.8518 & 0.2352 \\  
    
    \hline \hline
    12 & \multirow{10}{*}{\tabincell{c}{Token \\ level}} & \multirow{2}{*}{\tabincell{c}{Internal \\ embedding}} & Encoder embedding (Average) & 0.6956 & 0.3369 \\  
    13 &  & & Decoder embedding & \textbf{0.7991} & \textbf{0.2621} \\  \cline{3-3}
    14 &  & \multirow{6}{*}{Score} &Top-1 score logits & 0.7780 & 0.2791 \\  
    15 &  & &Top-5 score logits & 0.8524 & 0.2214 \\  
    16 &  & &Top-10 score logits & \textbf{0.8576} & \textbf{0.2195} \\  
    17 &  & &Top-20 score logits & 0.8517 & 0.2226 \\  
    18 &  & &All score logits & 0.8541 & 0.2234 \\  \cline{3-3} 
    19 &  &\multirow{3}{*}{Fusion} & ID (12) + ID (16) & 0.8380 & 0.2401 \\  
    20 &  & & ID (13) + ID (16) & \textbf{0.8787} & \textbf{0.2076} \\
    21 &  &  & ID (12) + ID (13) + ID (16) & 0.8251 & 0.2531 \\ \hline \hline
    
\end{tabular} 
}
\vspace{-0.5cm}
\end{table}

\begin{table*}[t]
\centering
\setlength{\abovecaptionskip}{0pt plus 1pt minus 4pt}
\caption{Speaker adapted performance (WER\%) of Conformer systems using LHUC, HUB, and PAct with or without SAT and Bayesian estimation on the Hub5'00, RT02 and RT03 test sets, before and after confidence score-based data selection and external transformer plus LSTM LMs rescoring. "O.V." represents "overall". "$\dagger$" and "$\ast$" denote a statistically significant WER difference over the baseline SI systems (sys. 1,26) and the corresponding non-Bayesian, point estimate-based adaptation baselines with no speaker data selection (sys. (2-4),(14-16),(27-29),(39-41)), respectively.}
\label{tab4:Vairous_form}
\resizebox{1.9\columnwidth}{!}{
\begin{tabular}{c|l|c|c|c|ccc|cccc|ccc} %
	\hline\hline
   \multirow{2}{*}{ID} & 
   \multirow{2}{*}{System} & 
   \multirow{2}{*}{\tabincell{c}{Adaptation \\ Parameters}} &
   \multirow{2}{*}{\tabincell{c}{Data \\ Selection}} &
   \multirow{2}{*}{\tabincell{c}{Language \\ Model}} &
   \multicolumn{3}{c|}{Hub5'00} & \multicolumn{4}{c|}{RT02} & 
    \multicolumn{3}{c}{RT03} 
    \\   \cline{6-15}
	&  & & & &  CHE & SWBD & O.V. & SWBD1 & SWBD2 & SWBD3 & O.V. & FSH & SWBD & O.V.   \\ \hline\hline

	1 & SI Model & - & - & \multirow{14}{*}{\xmark} & $15.0$ & $7.3$ & $11.1$ & $8.7$ & $12.9$ & $15.6$ & $12.6$ & $10.4$ & $16.4$ & $13.5$ \\ \cline{1-4} \cline{6-15}

	2 & LHUC & \multirow{3}{*}{Deterministic} &\multirow{6}{*}{\xmark}&  & $14.9$ & $7.3$ & $11.1$ & $8.7$ & $12.8$ & $15.5$ & $12.6$ & $10.4$ & $16.4$ & $13.5$ \\  
	3 & HUB & &&& $14.9$ & $7.3$ & $11.1$ & $8.8$ & $12.8$ & $15.4$ & $12.6$ & $10.4$ & $16.4$ & $13.5$ \\
	4 & PAct & &&  & $15.0$ & $7.2$ & $11.1$ & $8.7$ & $12.7$ & $15.8$ & $12.7$ & $10.4$ & $16.3$ & $13.5$\\

 	\cline{1-3}  \cline{6-15}

	5 & LHUC & \multirow{3}{*}{Bayesian} &&  & $14.8$ & $7.2$ & $11.0^{\dagger \ast}$ & $8.7$ & $12.8$ & $15.3^{\dagger}$& $12.5^{\dagger \ast}$ & $10.2^{\dagger \ast}$ & $16.3$ &$13.3^{\dagger \ast}$\\  
	6 & HUB & &&& $14.6^{\dagger}$ & $7.3$ & $11.0^{\dagger \ast}$ & $8.9$ & $12.7^{\dagger}$ & $15.3^{\dagger}$ & $12.5^{\dagger}$& $10.3$ & $16.1^{\dagger \ast}$ & $13.3^{\dagger \ast}$ \\
	7 & PAct & &&  & $14.6^{\dagger}$ & $7.2$ & $11.0^{\dagger \ast}$ & $8.7$ & $12.8$ & $15.6$ & $12.6$ & $10.4$ & $16.3$ & $13.5$ \\
    \cline{1-4} \cline{6-15}

	8 & LHUC & \multirow{3}{*}{Deterministic} &\multirow{6}{*}{\checkmark}&  & $14.5^{\dagger \ast}$ & $7.2$ & $10.8^{\dagger \ast}$ & $8.6$ & $12.7^{\dagger}$ & $15.2^{\dagger \ast}$& $12.4^{\dagger \ast}$ & $10.2^{\dagger \ast}$  & $16.1^{\dagger \ast}$ &$13.2^{\dagger \ast}$\\  
	9 & HUB & &&& $14.5^{\dagger \ast}$ & $7.1$ & $10.8^{\dagger \ast}$ & $8.6$ & $12.7^{\dagger}$ & $15.2^{\dagger}$ & $12.4^{\dagger \ast}$ & $\textbf {10.0}^{\dagger \ast}$ & $16.4$ & $13.3^{\dagger \ast}$ \\
	10 & PAct & &&  &$14.5^{\dagger \ast}$ & $7.1$ & $10.8^{\dagger \ast}$ & $8.7$ & $12.5^{\dagger}$ & $15.1^{\dagger \ast}$& $\textbf {12.3}^{\dagger \ast}$ & $10.2^{\dagger \ast}$  & $15.9^{\dagger \ast}$ & $13.1^{\dagger \ast}$\\
    \cline{1-3}  \cline{6-15}

	11 & LHUC & \multirow{3}{*}{Bayesian} && & $\textbf {14.2}^{\dagger \ast}$ & $\textbf {7.0}^{\dagger \ast}$ & $\textbf {10.6}^{\dagger \ast}$ & $8.6$ & $12.6^{\dagger \ast}$ & $\textbf {15.0}^{\dagger \ast}$& $\textbf {12.3}^{\dagger \ast}$ & $\textbf {10.0}^{\dagger \ast}$ &  $\textbf {15.8}^{\dagger \ast}$& $\textbf {13.0}^{\dagger \ast}$\\  
	12 & HUB & &&& ${14.3}^{\dagger \ast}$ & $\textbf {7.0}^{\dagger \ast}$ & ${10.7}^{\dagger \ast}$  & $8.6$ & $12.6^{\dagger \ast}$ & ${15.1}^{\dagger \ast}$& $\textbf {12.3}^{\dagger \ast}$& $\textbf {10.0}^{\dagger \ast}$ & $16.0^{\dagger \ast}$  & ${13.1 }^{\dagger \ast}$\\
	13 & PAct & &&  &$\textbf {14.2}^{\dagger \ast}$ & $\textbf {7.0}^{\dagger \ast}$ & $\textbf {10.6}^{\dagger \ast}$ & $\textbf {8.5}$ & $\textbf {12.3}^{\dagger \ast}$ & $15.2^{\dagger \ast}$& $\textbf {12.3}^{\dagger \ast}$ & $\textbf {10.0}^{\dagger \ast}$  & $15.9^{\dagger \ast}$ &$\textbf {13.0}^{\dagger \ast}$\\
    \hline 

	14 & LHUC-SAT & \multirow{4}{*}{Deterministic} &\multirow{7}{*}{\xmark}& \multirow{13}{*}{\xmark} & ${13.8}^{\dagger}$ & $7.1$ & ${10.5}^{\dagger}$ & $8.6$ & ${12.3}^{\dagger}$ & ${14.7}^{\dagger}$ & ${12.0}^{\dagger}$ & ${10.0}^{\dagger}$  & ${15.7}^{\dagger}$ &${12.9}^{\dagger}$\\  
	15 & HUB-SAT & &&& ${14.2}^{\dagger}$ & $7.1$ & ${10.7}^{\dagger}$ & $8.6$ & ${12.5}^{\dagger}$ & ${14.8}^{\dagger}$ & ${12.2}^{\dagger}$ & ${10.1}^{\dagger}$ &${15.7}^{\dagger}$ & ${13.0}^{\dagger}$\\
	16 & PAct-SAT & &&  & $14.7$ & $7.2$ & $ 11.0^{\dagger}$ & $8.6$ & $12.8$ & $15.6$ & $12.6$ & $10.3$ & $16.4$ & $13.5$ \\
	\cline{1-3}  \cline{6-15}
	
	17 & LHUC-SAT & \multirow{3}{*}{Bayesian} & && $13.7^{\dagger}$ & $6.9^{\dagger}$ & $10.3^{\dagger \ast}$ & $8.4^{\dagger}$ & $12.1^{\dagger \ast}$ & $14.3^{\dagger \ast}$ & $11.8^{\dagger \ast}$ & $9.8^{\dagger \ast}$ & $15.4^{\dagger \ast}$ & $12.7^{\dagger \ast}$\\  
	18 & HUB-SAT & &&&$13.7^{\dagger \ast}$ & $7.0^{\dagger}$ & $10.4^{\dagger \ast}$ & $8.5$ & $12.0^{\dagger \ast}$ & $14.2^{\dagger \ast}$ & $11.8^{\dagger \ast}$ & $9.7^{\dagger \ast}$ & $15.2^{\dagger \ast}$ & $12.5^{\dagger \ast}$\\
	19 & PAct-SAT && && $14.1^{\dagger \ast}$ & $7.0^{\dagger}$ & $10.6^{\dagger \ast}$ & $8.6$ & $12.5^{\dagger \ast}$ & $14.6^{\dagger \ast}$ & $12.1^{\dagger \ast}$ & $9.9^{\dagger \ast}$ & $15.4^{\dagger \ast}$ & $12.8^{\dagger \ast}$\\
	\cline{1-4} \cline{6-15}

	20 & LHUC-SAT & \multirow{3}{*}{Deterministic} &\multirow{6}{*}{\checkmark}&& ${13.7}^{\dagger}$ & ${6.9}^{\dagger}$  & ${10.3}^{\dagger \ast}$  & ${\textbf {8.3}^{\dagger \ast}}$  & ${12.0}^{\dagger \ast}$ & ${14.5}^{\dagger}$ & ${11.8}^{\dagger \ast}$ & ${9.8}^{\dagger \ast}$  & ${15.4 }^{\dagger \ast}$ & ${12.7}^{\dagger \ast}$ \\  
	21 & HUB-SAT &  & & & ${13.9}^{\dagger}$ & ${6.9}^{\dagger}$  & ${10.4}^{\dagger \ast}$  & ${8.4}^{\dagger}$  & ${12.1}^{\dagger \ast}$ & ${14.4}^{\dagger \ast}$ & ${11.8}^{\dagger \ast}$ & ${9.4}^{\dagger \ast}$  & ${15.4}^{\dagger \ast}$ & ${12.5}^{\dagger \ast}$ \\ 
	22 & PAct-SAT & & & & ${14.1}^{\dagger \ast}$ & $7.2$ & ${10.7}^{\dagger \ast}$ & $8.6$ & $12.7^{\dagger}$ & $15.2^{\dagger \ast}$ & ${12.3}^{\dagger \ast}$ & ${10.0}^{\dagger \ast}$ & ${15.7}^{\dagger \ast}$& ${13.0}^{\dagger \ast}$\\
	\cline{1-3} \cline{6-15}
	
	23 & LHUC-SAT & \multirow{3}{*}{Bayesian}&& &${\textbf {13.4}^{\dagger \ast}}$ & ${\textbf {6.8}^{\dagger \ast}}$ & ${\textbf {10.1}^{\dagger \ast}}$ &${\textbf {8.3}^{\dagger \ast}}$ & $\textbf {11.9}^{\dagger \ast}$ & ${14.1}^{\dagger \ast}$ & $\textbf {11.6}^{\dagger \ast}$ & ${9.5}^{\dagger \ast}$ & $\textbf {14.8}^{\dagger \ast}$ & $\textbf {12.3}^{\dagger \ast}$\\ 
	24 & HUB-SAT & &&& ${13.5}^{\dagger \ast}$ & ${6.9}^{\dagger}$ & ${10.2}^{\dagger \ast}$ & ${8.4}^{\dagger}$ & ${12.0}^{\dagger \ast}$ & $\textbf {13.8}^{\dagger \ast}$ & $\textbf {11.6}^{\dagger \ast}$ & $\textbf {9.4}^{\dagger \ast}$ & ${14.9}^{\dagger \ast}$ & $\textbf {12.3}^{\dagger \ast}$\\  
	25 & PAct-SAT & &&& ${13.6}^{\dagger \ast}$ & $7.2$ & ${10.4}^{\dagger \ast}$ & ${8.4}^{\dagger}$ & ${12.3}^{\dagger \ast}$ & ${14.6}^{\dagger \ast}$ & ${11.9}^{\dagger \ast}$ & ${9.8}^{\dagger \ast}$ & ${15.2}^{\dagger \ast}$ & ${12.6}^{\dagger \ast}$\\  
	\hline \hline
	
	26 & SI Model & - & -  &\multirow{13}{*}{\tabincell{c}{Transformer \\+ BiLSTM}} & $14.2$ & $6.8$  & $10.5$ & $8.4$ & $12.5$ & $14.3$ & $11.9$ & $9.4$ & $15.3$ & $12.4$ \\ \cline{1-4} \cline{6-15}

	27 &  LHUC & \multirow{3}{*}{Deterministic}  & \multirow{6}{*}{\xmark} && $14.2$ & $6.8$ & $10.5$ & $8.1^{\dagger}$ & $12.1^{\dagger}$ & $14.2$ & $11.7^{\dagger}$ & $9.4$ & $15.2$ & $12.4$ \\
	28 &  HUB &  &  & & $14.0$ & $6.8$ & $10.4^{\dagger}$ & $8.2$ & $12.1^{\dagger}$ & $14.1$ & $11.7^{\dagger}$ & $9.3$ & $15.3$ & $12.4$ \\
	29 &  PAct &  &  & & $14.2$ & $6.8$ & $10.5$ & $8.2$ & $12.0^{\dagger}$ & $14.3$ & $11.7^{\dagger}$ & $9.3$ & $15.3$ & $12.4$ \\
	\cline{1-3} \cline{6-15}

	30 &  LHUC & \multirow{3}{*}{Bayesian}  &  && $13.9$ & $6.7$ & $10.3^{\dagger \ast}$ & $8.0^{\dagger}$ & $11.7^{\dagger \ast}$ & $13.7^{\dagger \ast}$ & $11.4^{\dagger \ast}$ & $9.1^{\dagger \ast}$ & $15.0^{\dagger}$ & $12.1^{\dagger \ast}$ \\
	31 &  HUB &  &  & & $13.7^{\dagger}$ & $6.7$ & $10.2^{\dagger \ast}$ & $8.1^{\dagger}$ & $11.8^{\dagger \ast}$ & $14.0^{\dagger}$ & $11.5^{\dagger \ast}$ & $9.2^{\dagger}$ & $15.0^{\dagger}$ & $12.2^{\dagger \ast}$ \\
	32 &  PAct &  &  & & $13.8^{\dagger \ast}$ & $6.7$ & $10.3^{\dagger \ast}$ & $8.3$ & $11.8^{\dagger \ast}$ & $13.9^{\dagger \ast}$ & $11.5^{\dagger \ast}$ & $9.3$ & $14.8^{\dagger \ast}$ & $12.1^{\dagger \ast}$ \\
	\cline{1-4} \cline{6-15}
	
	33 &  LHUC & \multirow{3}{*}{Deterministic}  & \multirow{6}{*}{\checkmark} && $13.8^{\dagger \ast}$ & $6.6$ & $10.2^{\dagger \ast}$ & $\textbf {7.9}^{\dagger}$ & $11.5^{\dagger \ast}$ & $13.8^{\dagger \ast}$ & $11.3^{\dagger \ast}$ & $9.2$ & $14.8^{\dagger \ast}$ & $12.1^{\dagger \ast}$ \\
	34 &  HUB &  &  & & $13.3^{\dagger \ast}$ & $6.7$ & $\textbf {10.0}^{\dagger \ast}$ & $8.2$ & $11.7^{\dagger \ast}$ & $13.9^{\dagger}$ & $11.4^{\dagger \ast}$ & $9.1^{\dagger \ast}$ & $14.9^{\dagger \ast}$ & $12.1^{\dagger \ast}$\\
	35 &  PAct &  &  & & $13.6^{\dagger \ast}$ & $6.8$ & $10.2^{\dagger \ast}$ & $8.0^{\dagger}$ & $\textbf {11.3}^{\dagger \ast}$ & $13.7^{\dagger \ast}$ & $ 11.3^{\dagger \ast}$ & $9.1^{\dagger \ast}$ & $14.6^{\dagger \ast}$ & $12.0^{\dagger \ast}$ \\
	\cline{1-3} \cline{6-15}

	36 &  LHUC & \multirow{3}{*}{Bayesian}  & && $13.4^{\dagger \ast}$ & $\textbf {6.5}^{\dagger \ast}$ & $\textbf {10.0}^{\dagger \ast}$ & $\textbf {7.9}^{\dagger}$ & $11.5^{\dagger \ast}$ & $\textbf {13.5}^{\dagger \ast}$ & $\textbf {11.2}^{\dagger \ast}$ & $\textbf {9.0}^{\dagger \ast}$ & $\textbf {14.4}^{\dagger \ast}$ & $\textbf {11.8}^{\dagger \ast}$ \\
	37 &  HUB &  &  & &$\textbf {13.2}^{\dagger \ast}$ & $6.7$ & $\textbf {10.0}^{\dagger \ast}$& $8.1^{\dagger}$ & $11.5^{\dagger \ast}$ & $13.7^{\dagger \ast}$ & $11.3^{\dagger \ast}$ & $\textbf {9.0}^{\dagger \ast}$ & $14.7^{\dagger \ast}$ & $11.9^{\dagger \ast}$ \\
	38 &  PAct &  &  & & $13.5^{\dagger \ast}$ & $6.7$ & $10.1^{\dagger \ast}$ & $8.0^{\dagger}$ & $11.5^{\dagger \ast}$ & $\textbf {13.5}^{\dagger \ast}$ & $\textbf {11.2}^{\dagger \ast}$ & $\textbf {9.0}^{\dagger \ast}$ & $\textbf {14.4}^{\dagger \ast}$ & $\textbf {11.8}^{\dagger \ast}$ \\
	\hline

	39 &  LHUC-SAT & \multirow{3}{*}{Deterministic}  & \multirow{6}{*}{\xmark} &\multirow{13}{*}{\tabincell{c}{Transformer \\+ BiLSTM}} &$13.4^{\dagger}$ & $6.8$ & $10.1^{\dagger}$ & $8.2$ & $11.8^{\dagger}$ & $13.9^{\dagger}$ & $11.5^{\dagger}$ & $9.0^{\dagger}$  & $14.6^{\dagger}$ & $12.0^{\dagger}$ \\
	40 &  HUB-SAT &  &  & &$13.5^{\dagger}$ & $6.9$ & $10.2^{\dagger}$ & $8.1^{\dagger}$ & $11.7^{\dagger}$ & $13.8^{\dagger}$ & $11.4^{\dagger}$ & $9.0^{\dagger}$ & $14.8^{\dagger}$ & $12.1^{\dagger}$ \\
	41 &  PAct-SAT &  &  & & $14.0$ & $7.0$ & $10.5$ & $8.5$ & $12.1^{\dagger}$ & $14.4$ & $11.9$ & $9.4$ & $15.0$ & $12.3$ \\
	\cline{1-3} \cline{6-15}
	
	42 &  LHUC-SAT & \multirow{3}{*}{Bayesian}  &  &&$13.0^{\dagger \ast}$ & $6.7$ & $9.8^{\dagger \ast}$ & $8.1^{\dagger}$ & $11.6^{\dagger \ast}$ & $13.1^{\dagger \ast}$ & $11.1^{\dagger \ast}$ & $8.5^{\dagger \ast}$ & $14.0^{\dagger \ast}$ & $11.4^{\dagger \ast}$ \\
	43 &  HUB-SAT &  &  & &$13.0^{\dagger \ast}$ & $6.6^{\ast}$ & $9.8^{\dagger \ast}$ & $7.7^{\dagger \ast}$ & $11.3^{\dagger \ast}$ & $13.3^{\dagger \ast}$ & $11.0^{\dagger \ast}$ & $8.6^{\dagger \ast}$ & $14.2^{\dagger \ast}$ & $11.5^{\dagger \ast}$ \\
	44 &  PAct-SAT &  &  & & $13.4^{\dagger \ast}$ & $7.0$ & $10.2^{\dagger \ast}$ & $8.5$ & $11.9^{\dagger \ast}$ & $14.0^{\dagger \ast}$ & $11.6^{\dagger \ast}$ & $9.0^{\dagger \ast}$ & $14.6^{\dagger \ast}$ & $11.9^{\dagger \ast}$ \\
	\cline{1-4} \cline{6-15}

	45 &  LHUC-SAT &  \multirow{3}{*}{Deterministic}  & \multirow{6}{*}{\checkmark}& &${ {12.8}^{\dagger \ast}}$ & ${ {6.4}^{\dagger \ast}}$ & ${ {9.6}^{\dagger \ast}}$ & ${{7.9}^{\dagger \ast}}$& ${ {11.2}^{\dagger \ast}}$ & ${ {13.0}^{\dagger \ast}}$ & ${ {10.9}^{\dagger \ast}}$ & ${ {8.5}^{\dagger \ast}}$ & ${ {13.8}^{\dagger \ast}}$ & ${ {11.3}^{\dagger \ast}}$\\
	46 &  HUB-SAT &  &  & & $12.9^{\dagger \ast}$ & $6.5^{\dagger \ast}$ & $9.7^{\dagger \ast}$ & $7.9^{\dagger}$ & $\textbf {11.0}^{\dagger \ast}$ & $12.9^{\dagger \ast}$ & $10.8^{\dagger \ast}$ & $8.4^{\dagger \ast}$ & $14.0^{\dagger \ast}$ & $11.3^{\dagger \ast}$ \\
	47 &  PAct-SAT &  &  & &$13.5^{\dagger \ast}$ & $6.7^{\ast}$ & $10.1^{\dagger \ast}$ & $8.3$ & $11.7^{\dagger \ast}$ & $13.9^{\dagger \ast}$ & $11.5^{\dagger \ast}$ & $9.2^{\dagger \ast}$ & $14.5^{\dagger \ast}$ & $11.9^{\dagger \ast}$ \\
	\cline{1-3} \cline{6-15}

	48 &  LHUC-SAT & \multirow{3}{*}{Bayesian}   & & &${ {12.8}^{\dagger \ast}}$ & ${\textbf {6.2}^{\dagger \ast}}$ & ${\textbf {9.5}^{\dagger \ast}}$ & ${{7.7}^{\dagger \ast}}$& ${\textbf {11.0}^{\dagger \ast}}$ & ${\textbf {12.6}^{\dagger \ast}}$ & ${\textbf {10.6}^{\dagger \ast}}$ & ${\textbf {8.3}^{\dagger \ast}}$ & ${\textbf {13.6}^{\dagger \ast}}$ & ${\textbf {11.0}^{\dagger \ast}}$\\
	49 &  HUB-SAT &  &  & &$\textbf {12.6}^{\dagger \ast}$ & $6.3^{\dagger \ast}$ & $\textbf {9.5}^{\dagger \ast}$ & $\textbf {7.6}^{\dagger \ast}$ & $11.1^{\dagger \ast}$ & $\textbf {12.6}^{\dagger \ast}$ & $\textbf {10.6}^{\dagger}$ & $8.4^{\dagger \ast}$ & $\textbf {13.6}^{\dagger \ast}$ & $11.1^{\dagger \ast}$ \\
	50 &  PAct-SAT &  &  & &$12.9^{\dagger \ast}$ & $6.7^{\ast}$ & $9.8^{\dagger \ast}$ & $8.0^{\dagger \ast}$ & $11.4^{\dagger \ast}$ & $13.6^{\dagger \ast}$ & $11.2^{\dagger \ast}$ & $8.7^{\dagger \ast}$ & $14.0^{\dagger \ast}$ & $11.5^{\dagger \ast}$ \\
\hline \hline
\end{tabular}}
\vspace{-0.5cm}
\end{table*}

\subsubsection{Confidence Score Estimation for Adaptation} \label{sec63:cem}
Table \ref{tab31:predictorfeatures} shows the confidence score estimation performance of utterance-level and token-level CEMs using different prediction features. The equal error rate (EER) and area under the receiver operating characteristic (ROC) curve (AUC) are utilized as the evaluation metrics\footnote{EER is the operating point on the ROC curve where the false acceptance rate and false rejection rate are equal. AUC measures the average classification performance across the entire operating threshold range. The smaller the EER value or the larger the AUC value, the better performance.}. Several important trends can be observed. \textbf{a)} For both utterance and token level CEMs, the decoder embedding based prediction features (ID 2,13) consistently produced better AUC and EER scores than using the encoder embedding features (ID 1,12). \textbf{b)} For both CEMs, consistent improvements in AUC and EER were also obtained using n-best scores over 1-best scores (ID 4-6 vs. ID 3 or ID 15-18 vs. ID 14). As discussed in Sec. \ref{sec:51feature_cem}, the additional information from the competing hypotheses or tokens is beneficial for CEM prediction. \textbf{c)} The fusion of the decoder embedding with 10-best attention scores, or top-10 score logits (ID 10 or ID 20), produced the best performance and was selected as the prediction features for the utterance or token level CEMs training respectively in the following experiments. Such settings are selected based on their performance comparison against using various other features in Table~\ref{tab31:predictorfeatures}, e.g., internal Conformer embeddings\footnote{Both the Conformer encoder and decoder embeddings are considered as CEM prediction features in Table~\ref{tab31:predictorfeatures}.  First, the internal Conformer encoder embeddings are computed at an utterance level by averaging over the frame-level encoder embedding vectors obtained within each speech utterance. Hence, such utterance-level encoder embeddings are better synchronised with the utterance-level labels and their respective utterance-level confidence scores, than those at the individual output token level. This explains the performance of using utterance-level encoder embeddings to predict utterance-level confidence scores (ID 1, Table III) is better than predicting those at the token level (ID 12, Table III). However, both are clearly outperformed by using the fusion of the decoder embeddings with 10-best attention scores, or top-10 score logits (ID 10 or ID 20, Table III). These were therefore selected as the prediction features for the utterance or token level CEMs training respectively in the main experiments of this paper contained in Tables IV to VII. Second, the Conformer decoder embeddings are naturally synchronised with the output tokens on a one to one basis, thus no additional alignment between them is required (e.g., ID 13, Table III). Utterance decoder embeddings were produced by averaging over token-level decoder embeddings within each utterance (e.g., ID 2, Table III).}.

As further shown in Fig.~\ref{fig:cem}, the effectiveness of the proposed confidence score-based speaker-level data selection is evaluated through an extensive comparison with various selection approaches using different confidence measures on the Hub5’00, RT02 and RT03 test sets. The WER performance of adapted LHUC-SAT Conformer systems with varying amounts of selected speaker data is illustrated. Several trends can be observed. \textbf{a)} Across varying amounts of adaptation data on all three test sets, the adaptation performance obtained by the proposed utterance-level and token-level CEM smoothed Conformer attention probabilities (blue and orange lines) consistently outperformed those using the comparable baseline selection methods using un-mapped, original attention probabilities (green lines) and their further combination with CTC probabilities (purple lines). \textbf{b)} When Bayesian learning was performed to avoid the risk of over-fitting, the performance of the Bayesian adaptation using token-level CEM smoothed confidence measures (orange dotted lines) was comparable to that using the oracle, ground truth WER-based selection approach (red dotted line at the bottom in all three figures). In addition, the token-level CEM (orange dotted lines) produced slightly better Bayesian adaptation performance than the utterance-level CEM (blue dotted lines). This may be attributed to the fact that the fine-grained token-level prediction features are more informative in distinguishing correct hypotheses from erroneous ones than utterance-level features that are obtained by averaging over token-level features. For the best performing token-level CEM smoothed confidence score-based data selection (orange dotted lines), the best operating data selection percentile point is around 80\%, which is adopted in the following adaptation experiments in Sec. \ref{sec64:main}.

\subsubsection{Performance of Confidence Score Based Bayesian Adaptation}\label{sec64:main} The adapted performance of the Conformer systems using LHUC, HUB, and PAct evaluated on all three test sets is further shown in Table \ref{tab4:Vairous_form} and Table\ref{tab:900hrswbd}, several main trends can be observed. 

\textbf{a)} By using the CEM smoothed confidence scores in speaker data selection for adaptation, the resulting LHUC-SAT, HUB-SAT and PAct-SAT speaker adapted Conformer systems (sys. 20-22) consistently outperformed the baseline SI Conformer system (sys. 1) and the corresponding baseline adaptation methods without data selection (sys. 14-16) across all three test sets. In particular, the confidence score-based HUB-SAT Conformer system (sys. 21) produced statistically significant WER reductions of \textbf{1.0\%} absolute (\textbf{9.6\%} relative) and \textbf{0.7\%} absolute (\textbf{6.9\%} relative) on the FSH subset of RT03 test set over the baseline SI system (sys. 1) and the corresponding adapted system without data selection (sys. 15), respectively. 

\textbf{b)} When Bayesian learning was further applied to the confidence score-based adaptation to address the increased data sparsity caused by data selection, additional performance improvements were consistently obtained by the Bayesian adapted LHUC-SAT, HUB-SAT, PAct-SAT Conformer systems (sys. 23-25) over that of the comparable non-Bayesian deterministic estimate-based adaptation methods (sys. 20-22) on all three test sets. For example, a statistically significant WER reduction of \textbf{0.6\%} absolute (\textbf{4.2\%} relative) was obtained by the Bayesian adapted HUB-SAT Conformer system (sys. 24) over the non-Bayesian adapted HUB-SAT Conformer system (sys. 21) on the SWBD3 subset of RT02 data. 

\textbf{c)} Similar performance improvements were obtained when only test-time adaptation was performed. The LHUC, HUB or PAct adapted Conformer systems (sys. 11-13) using both confidence score-based data selection and Bayesian learning consistently outperformed the corresponding standard adaptation (sys. 2-4) without data selection, and the confidence score based non-Bayesian adapted Conformer systems (sys. 8-10). 

\textbf{d)} When the external Transformer plus Bidirectional LSTM language models were further used in n-best outputs rescoring, similar consistent performance improvements were observed. The best adapted performance was produced by the proposed confidence score-based Bayesian adapted LHUC-SAT Conformer system (sys. 48). Overall statistically significant WER reductions of \textbf{1.0\%}, \textbf{1.3\%} and \textbf{1.4\%} absolute (\textbf{9.5\%}, \textbf{10.9\%} and \textbf{11.3\%} relative) were obtained on the Hub5’00, RT02 and RT03 test sets respectively over the baseline SI Conformer system (sys. 26). 

\textbf{e)} As shown in Table \ref{tab:900hrswbd}, by further incorporating speed perturbation-based data augmentation in system training, the trends found in Table IV were still retained. Absolute WER reductions of \textbf{1.0\%-1.3\%} and \textbf{0.7\%-1.0\%} were obtained over the baseline SI system and the corresponding non-Bayesian, point estimate-based adaptation (sys. 8 vs. sys. 5,6) across all three test sets respectively after external LM rescoring. The experimental results shown in Table \ref{tab4:Vairous_form} and \ref{tab:900hrswbd} indicated that the proposed confidence score-based adaptation, speed perturbation and external language model rescoring are complementary, with their improvements being largely additive. 

The best performance of the proposed confidence score-based Bayesian adapted LHUC-SAT Conformer system (sys. 8 in Table \ref{tab:900hrswbd}) is further contrasted in Table \ref{tab:table_comparison} with those of the state-of-the-art performance achieved on the 300-hour Switchboard task using the hybrid and E2E ASR systems reported in the latest literature to demonstrate its competitiveness. 

\begin{table*}[!t]
\centering
\setlength{\abovecaptionskip}{0pt plus 1pt minus 4pt}
\caption{Performance (WER\%) of data augmented 900-hour trained LHUC-SAT Conformer systems with or without Bayesian speaker adaptation on all three test sets, before and after confidence score-based data selection and external Transformer plus LSTM LMs rescoring. "$\dagger$" and "$\ast$" denote a statistically significant WER difference over the corresponding baseline SI systems (sys. 1,5) and the corresponding non-Bayesian, point estimate-based adaptations without data selection (sys. 2,6), respectively.}
\label{tab:900hrswbd}
\resizebox{1.8\columnwidth}{!}{
\begin{tabular}{c|l|c|c|c|ccc|cccc|ccc} %
	\hline\hline
   \multirow{2}{*}{ID} & 
   \multirow{2}{*}{System} & 
   \multirow{2}{*}{\tabincell{c}{Adaptation \\ Parameters}} &
   \multirow{2}{*}{\tabincell{c}{Data \\ Selection}} &
   \multirow{2}{*}{\tabincell{c}{Language \\ Model}} &
   \multicolumn{3}{c|}{Hub5'00} & \multicolumn{4}{c|}{RT02} &
   \multicolumn{3}{c}{RT03} 
    \\   \cline{6-15}
	&  & & & &  CHE & SWBD & O.V. & SWBD1 & SWBD2 & SWBD3 & O.V. & FSH & SWBD & O.V.   \\ \hline\hline

	1 & SI Model & - & - & \multirow{4}{*}{\xmark} & $15.1$ & $7.2$ & $11.1$ & $8.5$ & $12.6$ & $14.7$ & $12.1$ & $10.3$ & $16.1$ & $13.3$ \\
	\cline{1-4}
	2 & \multirow{3}{*}{LHUC-SAT} & Deterministic & \xmark &  & $13.9^{\dagger}$ & $7.0$ & $10.5^{\dagger}$ & $8.2^{\dagger}$ & $12.1^{\dagger}$ & $14.1^{\dagger}$ & $11.7^{\dagger}$ & $9.9^{\dagger}$ & $15.7^{\dagger}$ & $12.9^{\dagger}$ \\
	3 &  & Deterministic & \checkmark&  & $13.5^{\dagger \ast}$ & $\textbf {6.9}^{\dagger}$ & $10.2^{\dagger \ast}$ & $8.0^{\dagger}$ & $11.9^{\dagger \ast}$ & $13.4^{\dagger \ast}$ & $11.2^{\dagger \ast}$ & $9.4^{\dagger \ast}$ & $15.0^{\dagger \ast}$ & $12.3^{\dagger \ast}$ \\
	4 &  & Bayesian & \checkmark &  & $\textbf {13.0}^{\dagger \ast}$ & $\textbf {6.9}^{\dagger}$ & $\textbf {10.0}^{\dagger \ast}$ & $\textbf {7.9}^{\dagger \ast}$ & $\textbf {11.8}^{\dagger \ast}$ & $\textbf {13.2}^{\dagger \ast}$ & $\textbf {11.1}^{\dagger \ast}$ & $\textbf {9.3}^{\dagger \ast}$ & $\textbf {14.7}^{\dagger \ast}$ & $\textbf {12.1}^{\dagger \ast}$ \\
	\hline \hline
 
	5 & SI Model & - & - & \multirow{4}{*}{\tabincell{c}{Transformer \\+ BiLSTM}} & $14.0$ & $6.8$ & $10.4$ & $8.0$ & $12.0$ & $13.4$ & $11.3$ & $9.3$ & $15.1$ & $12.3$  \\
	\cline{1-4}
	6 & \multirow{3}{*}{LHUC-SAT} & Deterministic & \xmark &  & $13.5^{\dagger}$ & $6.7$ & $10.1^{\dagger}$ & $7.8$ & $11.4^{\dagger}$ & $12.6^{\dagger}$ & $10.8^{\dagger}$ & $8.9^{\dagger}$ & $14.7^{\dagger}$ & $12.0^{\dagger}$ \\
	7 &  & Deterministic & \checkmark&  & $12.5^{\dagger \ast}$ & $6.6$ & $9.6^{\dagger \ast}$ & $7.6^{\dagger}$ & $\textbf{10.7}^{\dagger \ast}$ & $12.0^{\dagger \ast}$ & $10.3^{\dagger \ast}$ & $8.3^{\dagger \ast}$ & $13.8^{\dagger \ast}$ & $11.2^{\dagger \ast}$ \\
	8 &  & Bayesian & \checkmark &  & $\textbf {12.4}^{\dagger \ast}$ & $\textbf {6.5}^{\dagger}$ & $\textbf {9.4}^{\dagger \ast}$ & $\textbf {7.4}^{\dagger \ast}$ & $\textbf {10.7}^{\dagger \ast}$ & $\textbf {11.8}^{\dagger \ast}$ & $\textbf {10.1}^{\dagger \ast}$ & $\textbf {8.2}^{\dagger \ast}$ & $\textbf {13.6}^{\dagger \ast}$ & $\textbf {11.0}^{\dagger \ast}$ \\
	\hline\hline
\end{tabular}}
\vspace{-0.5cm}
\end{table*}
\begin{table*}[htbp]
\centering
\setlength{\abovecaptionskip}{0pt plus 1pt minus 4pt}
\caption{Performance (WER\%) contrast between the best confidence score-based Bayesian LHUC-SAT Conformer system and other state-of-the-art systems on the Hub5’00 and RT03 data.}
\label{tab:table_comparison}
\resizebox{1.5\columnwidth}{!}{
\begin{tabular}{c|l|l|ccc|ccc} %
	\hline\hline
    \multirow{2}{*}{ID} & 
    \multirow{2}{*}{System}  &
    \multirow{2}{*}{\#Para.} &
    \multicolumn{3}{c|}{Hub5'00 } & 
    \multicolumn{3}{c}{RT03} 
    \\   \cline{4-9}
	& &  & CHE & SW. & O.V. & FSH & SW. & O.V.  \\ \hline\hline
	1 & RWTH-2019 Hybrid\cite{Kitza2019CumulativeAF} & - & 13.5 & 6.7 & 10.2 & - & - & - \\ \hline 
	2 & Google-2019 LAS\cite{Park2019SpecAugmentAS} & -  & 14.1 & 6.8 & (10.5) & - & - & - \\  \hline 
	3 & \multirow{3}{*}{IBM-2020 AED \cite{tuske2020single}}  & 29M & 14.6 & 7.4 & (11.0) & - & - & - \\
	4 &   & 75M  & 13.4 & 6.8 & (10.1) & - & - & - \\
	5 &   & 280M & 12.5 & 6.4 & 9.5 & 8.4 & 14.8 & (11.7) \\ \hline 
	6 & Salesforce-2020 Transformer\cite{wang2020investigation} & -  & 13.3 & 6.3 & (9.8) & - &- & 11.4 \\ \hline 
	7 & CUHK-2021 BLHUC-Hybrid\cite{Xie2021BayesianLF} & 15.2M &  12.7 & 6.7 & 9.7 & 7.9 & 13.4 & 10.7 \\ \hline 
	8 & IBM-2021 CFM-AED\cite{tuske2021limit} & 68M  & 11.2 & 5.5 & (8.4) & 7.0 & 12.6 & (9.9) \\ \hline
	9 & IBM-2021 RNN-T\cite{saon2021advancing} & 57M  & 13.1 & 6.3 & (9.7) & 9.5 & 15.4 & (12.6) \\ \hline
	10 & RWTH-2022 CFM-Hybrid\cite{zeineldeen2022conformer} & - & 14.1 & 6.3 & 10.3 & - & - & - \\ \hline
	11 & RWTH-2022 adapted CFM-Hybrid \cite{zeineldeen2022improving} & - & 12.1 & 6.3 & 9.2 & - & - & - \\ \hline
	12 & \multirow{2}{*}{\tabincell{l}{\textbf{SI Conformer \{TAB. \ref{tab:900hrswbd}-sys. 5\} (Ours)}\\ \textbf{+ Adaptation \{TAB. \ref{tab:900hrswbd}-sys. 8\} (Ours)}}} & 45M & 14.0 & 6.8  & 10.4 & 9.3 & 15.1 & 12.3 \\
	13 & & + 30k & 12.4 & 6.5 & 9.4 & 8.2 & 13.6 & 11.0  \\\hline \hline
\end{tabular} }
\vspace{-0.2cm}
\end{table*}

\begin{table*}[!t]
\centering
\setlength{\abovecaptionskip}{0pt plus 1pt minus 6pt}
\caption{Performance (WER\%) of LHUC and LHUC-SAT Conformer systems with or without Bayesian estimation and confidence score-based data selection on the AMI development and evaluation sets of ihm, sdm and mdm conditions, before and after external Transformer rescoring. $\dagger$ and $\ast$ denote a statistically significant WER difference over the corresponding baseline SI systems (sys. 1,10) and the corresponding point estimate-based adaptations without data selection (sys. 2,6,11,15), respectively.}
\label{tab:ami_main}
\resizebox{1.9\columnwidth}{!}{
\begin{tabular}{c|c|c|c|c|cccc|cccc} %
	\hline \hline
    \multirow{2}{*}{ID} & 
    \multirow{2}{*}{\tabincell{c}{System}}  &
    \multirow{2}{*}{\tabincell{c}{Adaptation \\Parameters}} &
    \multirow{2}{*}{\tabincell{c}{Data\\Selection}} & 
    \multirow{2}{*}{\tabincell{c}{Language\\Model}} &   \multicolumn{4}{c|}{DEV } & \multicolumn{4}{c}{EVAL}
    \\   \cline{6-13}
	& & &  & & IHM & SDM & MDM & O.V. & IHM & SDM & MDM & O.V. \\ \hline\hline

	1  & SI Model & - & - & \multirow{9}{*}{\xmark}& $20.9$ & $38.0$ & $34.9$ & $31.3$ & $20.7$ & $42.1$ & $37.8$ & $33.6$ \\ \cline{1-4} \cline{6-13}
	2  & \multirow{4}{*}{LHUC} & Deterministic & \multirow{2}{*}{\xmark} & & $20.7^{\dagger}$ & $38.0$ & $34.9$ & $31.3$ & $20.5^{\dagger}$ & $41.6^{\dagger}$ & $37.8$ & $33.3^{\dagger}$ \\ 
	3  &  & Bayesian &  & & $20.5^{\dagger \ast}$ & $37.7^{\dagger \ast}$ & $34.7$ & $30.9^{\dagger \ast}$ & $20.3^{\dagger \ast}$ & $41.3^{\dagger}$ & $37.5^{\dagger \ast}$ & $33.0^{\dagger \ast}$ \\ \cline{3-4}
	4   &  & Deterministic & \multirow{2}{*}{\checkmark} & & $20.3^{\dagger \ast}$ & $37.4^{\dagger \ast}$ & $34.4^{\dagger \ast}$ & $30.7^{\dagger \ast}$ & $19.9^{\dagger \ast}$ & $41.0^{\dagger \ast}$ &$37.1^{\dagger \ast}$ & $32.7^{\dagger \ast}$ \\ 
	5   & & Bayesian &  & & $\textbf{20.1}^{\dagger \ast}$ & $\textbf {37.2}^{\dagger \ast}$ & $\textbf {34.2}^{\dagger \ast}$ & $\textbf{30.5}^{\dagger \ast}$ & $\textbf{19.7}^{\dagger \ast}$ &  $\textbf {40.9}^{\dagger \ast}$ &  $\textbf {36.7}^{\dagger \ast}$ & $\textbf{32.4}^{\dagger \ast}$\\
	
	\cline{2-4} \cline{6-13}

	6   & \multirow{4}{*}{LHUC-SAT} & Deterministic & \multirow{2}{*}{\xmark} & & $20.4^{\dagger}$ & $36.4^{\dagger}$ & $33.8^{\dagger}$ & $30.2^{\dagger}$ & $20.1^{\dagger}$ & $41.2^{\dagger}$ & $35.9^{\dagger}$ & $32.4^{\dagger}$\\ 
	7  &  & Bayesian &  & & $19.9^{\dagger \ast}$ & $36.0^{\dagger \ast}$ & $33.0^{\dagger \ast}$ & $29.6^{\dagger \ast}$ & $19.7^{\dagger \ast}$ & $39.6^{\dagger \ast}$ &  $35.6^{\dagger \ast}$ & $31.6^{\dagger \ast}$\\ \cline{3-4}
	8   &  & Deterministic & \multirow{2}{*}{\checkmark} & & $19.7^{\dagger \ast}$ & $35.1^{\dagger \ast}$ & $32.8^{\dagger \ast}$ & $29.2^{\dagger \ast}$ & $19.1^{\dagger \ast}$ & $38.6^{\dagger \ast}$ & $34.8^{\dagger \ast}$ & $30.8^{\dagger \ast}$\\ 
	9   & & Bayesian &  & & $\textbf {19.4}^{\dagger \ast}$ & $\textbf {34.6}^{\dagger \ast}$ & $\textbf {32.2}^{\dagger \ast}$ & $\textbf{28.7}^{\dagger \ast}$ & $\textbf{18.8}^{\dagger \ast}$ & $\textbf {37.9}^{\dagger \ast}$ & $\textbf {34.0}^{\dagger \ast}$ & $\textbf{30.2}^{\dagger \ast}$\\ 
	
	\hline \hline

	10  & SI Model & - & - & \multirow{9}{*}{Transformer}& $20.2$ & $36.9$ & $33.9$ & $30.4$ & $19.8$ & $40.7$ & $36.6$ & $32.4$\\ \cline{1-4} \cline{6-13}
	11   & \multirow{4}{*}{LHUC} & Deterministic & \multirow{2}{*}{\xmark} & & $20.1$ & $36.9$ & $33.8$ & $30.3$ & $19.8$ & $40.6$ & $36.6$ & $32.4$\\ 
	12  &  & Bayesian &  & & $19.9^{\dagger \ast}$ & $36.7$ & $33.5^{\dagger \ast}$ & $30.0^{\dagger \ast}$ & $19.7$ & $40.2^{\dagger \ast}$ & $36.0^{\dagger \ast}$ & $32.0^{\dagger \ast}$\\ \cline{3-4}
	13   &  & Deterministic & \multirow{2}{*}{\checkmark} & & $19.5^{\dagger \ast}$ & $36.2^{\dagger \ast}$ & $33.3^{\dagger \ast}$ & $29.7^{\dagger \ast}$ & $19.1^{\dagger \ast}$ & $39.7^{\dagger \ast}$ & $35.7^{\dagger \ast}$ & $31.5^{\dagger \ast}$ \\ 
	14   & & Bayesian &  & & $\textbf {19.3}^{\dagger \ast}$ & $\textbf {35.9}^{\dagger \ast}$ & $\textbf {33.0}^{\dagger \ast}$ & $\textbf{29.4}^{\dagger \ast}$ & $\textbf {19.0}^{\dagger \ast}$ & $\textbf {39.5}^{\dagger \ast}$ & $\textbf {35.4}^{\dagger \ast}$ & $\textbf{31.3}^{\dagger \ast}$\\ \cline{2-4} \cline{6-13}
	
	15   & \multirow{4}{*}{LHUC-SAT} & Deterministic & \multirow{2}{*}{\xmark} & & $19.7^{\dagger}$ & $35.3^{\dagger}$ & $32.7^{\dagger}$ & $29.3^{\dagger}$ & $19.1^{\dagger}$ & $38.8^{\dagger}$ & $34.9^{\dagger}$ & $30.9^{\dagger}$\\ 
	16  &  & Bayesian &  & & $19.4^{\dagger \ast}$ & $34.5^{\dagger \ast}$ & $31.8^{\dagger \ast}$ & $28.6^{\dagger \ast}$ & $18.7^{\dagger \ast}$ & $38.1^{\dagger \ast}$ & $34.0^{\dagger \ast}$ & $30.3^{\dagger \ast}$\\ \cline{3-4}
	17   &  & Deterministic & \multirow{2}{*}{\checkmark} & & $19.1^{\dagger \ast}$ & $34.0^{\dagger \ast}$ & $31.3^{\dagger \ast}$ & $28.2^{\dagger \ast}$ & $18.3^{\dagger \ast}$ & $37.3^{\dagger \ast}$ & $33.4^{\dagger \ast}$ & $29.7^{\dagger \ast}$\\ 
	18   & & Bayesian &  & & $\textbf {18.8}^{\dagger \ast}$ & $\textbf {33.5}^{\dagger \ast}$ & $\textbf {30.9}^{\dagger \ast}$ & $\textbf{27.7}^{\dagger \ast}$ & $\textbf {18.0}^{\dagger \ast}$ & $\textbf {36.6}^{\dagger \ast}$ & $\textbf {32.9}^{\dagger \ast}$ & $\textbf{29.1}^{\dagger \ast}$\\
	\hline \hline
	
\end{tabular} 
}
\end{table*}
\subsection{Experiments on AMI Data} \label{sec6:ami}
\subsubsection{Experimental Setup} \label{sec6b1:setup}
The speed perturbed AMI speech data with nearly 233-hour audio collected from 1641 speakers using both headset microphone and distant microphone arrays from the meeting environment was used for training. The 8.9-hour development set (DEV) with 72 speakers and the 8.6-hour evaluation set (EVAL) with 63 speakers recorded under the individual headset microphone (ihm), single distant microphone (sdm) and multiple distant microphone (mdm) conditions were used for evaluation. The ESPnet Conformer ASR system configurations were the same as those for the Switchboard task presented in Sec. \ref{sec61:setup}, except that the BPE token of size 100 was employed. The external Transformer LM trained on the AMI transcripts was used for further rescoring. 

\subsubsection{Performance of Confidence Score Based LHUC Adaptation}\label{sec6b2:main} The experimental results previously shown in Table \ref{tab4:Vairous_form} suggested that LHUC adaptation achieved the best performance compared to other adaptation methods on the Switchboard task. Therefore, a series of comparable LHUC adaptation experiments were conducted on the AMI data to further demonstrate the efficacy and advantages of the proposed confidence score-based adaptation. 

\begin{figure}[htbp]
    \centering
    \setlength{\abovecaptionskip}{0pt plus 1pt minus 4pt}
    \includegraphics[scale=0.5]{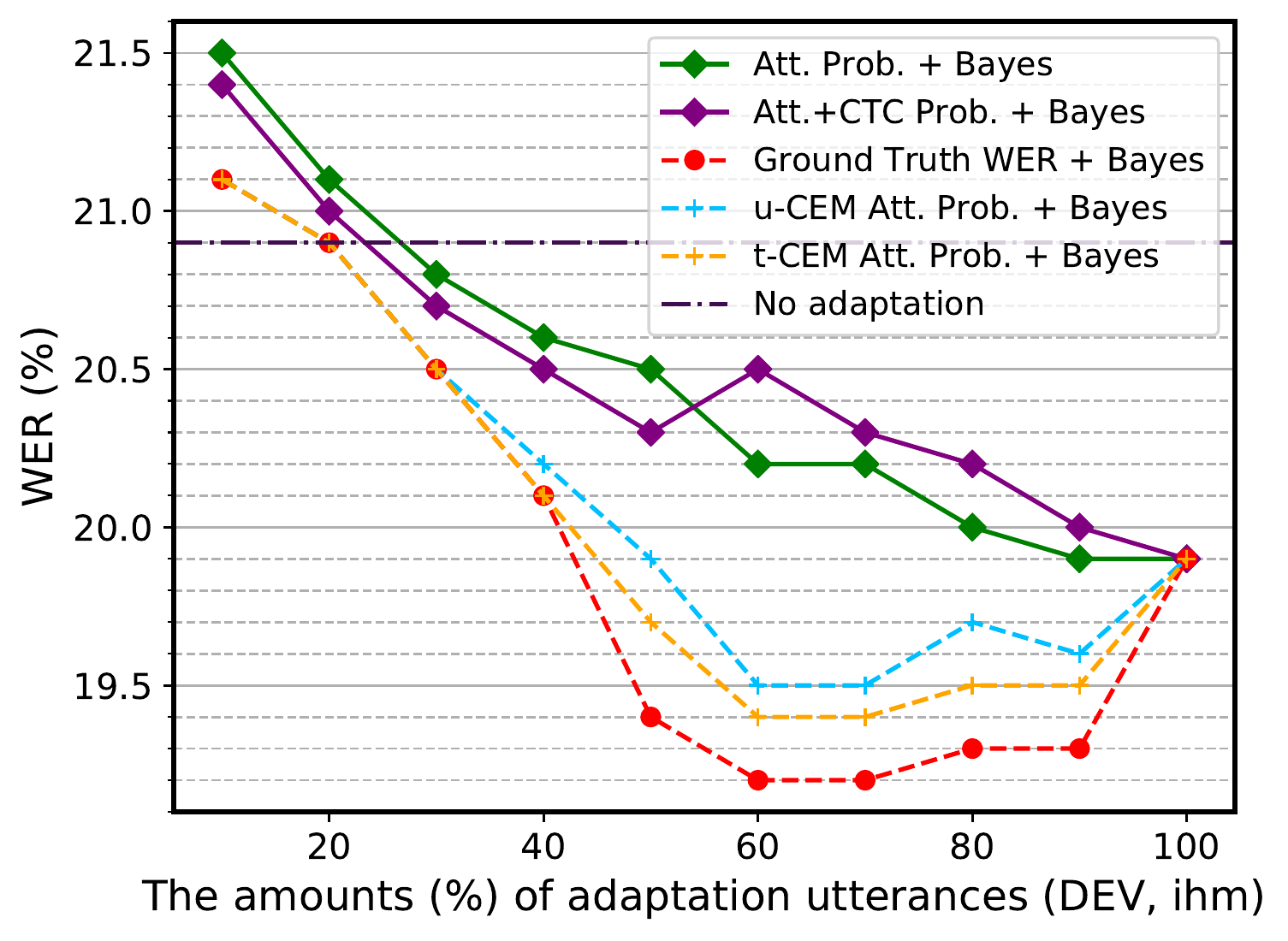}
    \caption{Performance (WER\%) of adapted LHUC-SAT Conformer systems with varying amounts of adaptation data selected using different confidence measures. All naming conventions follow Fig. \ref{fig:cem}.}
    \label{fig:ami_dev1}
\end{figure}
Fig.~\ref{fig:ami_dev1} shows the adapted performance of LHUC-SAT Conformer systems using various speaker data selection methods evaluated on the ihm development data. A general trend similar to that found in Fig.~\ref{fig:cem} can be observed across various selected subsets. The proposed utterance-level (blue dotted line) and token-level (orange dotted line) CEMs consistently produced better adaptation performance than the comparable baseline selection using the original Conformer output attention or attention plus CTC probabilities (green and purple lines). The performance of the proposed token-level CEM (orange dotted line) is the closest to that using the oracle, ground truth WER-based data selection (red dotted line). The best percentile operating point for subset selection is approximately 70\%. 

Using the above top 70\% selected adaptation data based on the token-level CEM, the performance of the proposed confidence score-based LHUC adaptation evaluated on the DEV and EVAL sets under the ihm, sdm and mdm conditions were shown in Table VII. Several main trends can be observed. 

\textbf{a)} The confidence score-based adapted LHUC-SAT Conformer system (sys. 8) consistently produced statistically significant WER reductions of \textbf{0.7\%} (DEV, ihm) to \textbf{2.6\%} (EVAL, sdm) absolute (\textbf{3.4\%} to \textbf{6.3\%} relative) over the corresponding baseline adaptation (sys. 6) without using data selection. 

\textbf{b)} By further incorporating Bayesian estimation in test-time adaptation, additional consistent performance improvements of up to \textbf{0.8\%} (EVAL, mdm) absolute WER reduction were obtained by the confidence score-based Bayesian adapted Conformer system (sys. 9) over that using non-Bayesian, point estimate-based adaptation (sys. 8) on both test sets. 

\textbf{c)} Similar trends were also found when performing test-time adaptation of non-SAT, SI Conformer systems. The proposed confidence score-based Bayesian LHUC adapted Conformer system (sys. 5) consistently outperformed the corresponding baseline adaptation without data selection (sys. 2) and confidence score-based non-Bayesian adapted system (sys. 4). 

\textbf{d)} Consistent performance improvements were retained after n-best rescoring using the external Transformer language model. The best performance was produced by the confidence score-based Bayesian adapted LHUC-SAT system (sys. 18). Overall statistically significant WER reductions of \textbf{2.7\%} and \textbf{3.3\%} absolute (\textbf{8.9\%} and \textbf{10.2\%} relative) were obtained over the baseline SI Conformer system (sys. 10) on the DEV and EVAL sets, respectively.

\section{Conclusion} \label{sec:7}
This paper proposed confidence score-based unsupervised Bayesian speaker adaptation approaches using a variety of compact and data efficient speaker-dependent (SD) parameter representations to facilitate both speaker adaptive training and test-time unsupervised speaker adaptation of state-of-the-art Conformer ASR systems. The sensitivity of adaptation performance to supervision quality was reduced using speaker-level data selection based on the confidence scores produced by specially designed confidence estimation modules. The data sparsity issue further increased by the confidence score-based speaker data selection was mitigated using Bayesian learning. Experiments on the benchmark 300-hour Switchboard corpus and the 233-hour AMI dataset suggested that the proposed confidence score-based adaptation scheme consistently outperformed the baseline speaker-independent Conformer model and conventional non-Bayesian, point estimate-based adaptation using no speaker data selection. Similar consistent performance improvements were retained after external language model rescoring. Further research will focus on rapid on-the-fly adaptation for Conformer and other E2E ASR systems. 

\section*{Acknowledgment}
This research is supported by Hong Kong RGC GRF grant No. 14200021, 14200220, Innovation \& Technology Fund grant No. ITS/254/19 and ITS/218/21, National Key R\&D Program of China (2020YFC2004100), the National Natural Science Foundation of China (NSFC) Grant 62106255, the Guangzhou Civil Affairs Science and Technology Foundation Grant 2022MZK02, and the Open Research Fund of Guangxi Key Lab of Human-machine Interaction and Intelligent Decision GXHIID2202.

\bibliography{main.bib}
\bibliographystyle{IEEEtran.bst}

\end{document}